\documentclass[twocolumn,amsmath]{aastex702}

\accepted{for publication in ApJ on November 16, 2026}

\begin{document}

\title{Tracing Accretion and Inner-Disk Gas Evolution in the Orion Nebula via Br$\alpha$ and CO}

\correspondingauthor{Konstantin Getman}
\email{kug1@psu.edu}

\author[0000-0002-6137-8280]{Konstantin V. Getman}
\affiliation{Department of Astronomy \& Astrophysics \\
Pennsylvania State University \\ 
525 Davey Laboratory \\
University Park, PA 16802, USA}
\email{kug1@psu.edu}

\author[0000-0001-9698-4080]{Mar\'ia Claudia Ram\'irez-Tannus}
\affiliation{Max-Planck-Institut f\"ur Astronomie\\
K\"onigstuhl 17\\
D-69117 Heidelberg, Germany}
\email{ramirez@mpia.de}

\author[0009-0003-7663-5280]{Jenny Frediani}
\affiliation{Department of Astronomy, Stockholm University\\
AlbaNova University Centre\\
106 91 Stockholm, Sweden}
\email{jenny.frediani@astro.su.se}

\author[0000-0003-2822-2951]{Kevin L. Luhman}
\affiliation{Department of Astronomy \& Astrophysics \\
Pennsylvania State University \\ 
525 Davey Laboratory \\
University Park, PA 16802, USA}
\affiliation{Center for Exoplanets and Habitable Worlds \\ 525 Davey Laboratory \\ The Pennsylvania State University \\ University Park, PA 16802, USA}
\email{kll207@psu.edu}

\author[0000-0002-5077-6734]{Eric D. Feigelson}
\affiliation{Department of Astronomy \& Astrophysics \\
Pennsylvania State University \\ 
525 Davey Laboratory \\
University Park, PA 16802, USA}
\affiliation{Center for Exoplanets and Habitable Worlds \\ 525 Davey Laboratory \\ The Pennsylvania State University \\ University Park, PA 16802, USA}
\email{e5f@psu.edu}

\author[0000-0002-6278-9006]{Bayron Portilla-Revelo}
\affiliation{Department of Astronomy \& Astrophysics \\
Pennsylvania State University \\ 
525 Davey Laboratory \\
University Park, PA 16802, USA}
\affiliation{Center for Exoplanets and Habitable Worlds \\ 525 Davey Laboratory \\ The Pennsylvania State University \\ University Park, PA 16802, USA}
\email{bmp5924@psu.edu}

\author[0000-0003-1817-6576]{Christian Rab}
\affiliation{University Observatory, Faculty of Physics\\
Ludwig-Maximilians-Universit\"at M\"unchen\\
Scheinerstr. 1\\
D-81679 Munich, Germany}
\affiliation{Max-Planck-Institut f\"ur extraterrestrische Physik\\
Giessenbachstrasse 1\\
D-85748 Garching, Germany}
\email{rab@mpe.mpg.de}

\author[0009-0002-5380-403X]{Valentin Brunn}
\affiliation{INAF--Osservatorio Astrofisico di Arcetri\\
Largo E. Fermi 5\\
50125 Firenze, Italy}
\email{valentin.brunn@inaf.it}

\author[0000-0001-8694-4966]{Jan Forbrich}
\affiliation{Centre for Astrophysics Research\\
Department of Physics, Astronomy and Mathematics\\
University of Hertfordshire\\
Hatfield, AL10 9UW, UK}
\email{j.forbrich@herts.ac.uk}

\author[0000-0002-9593-7618]{Thomas J. Haworth}
\affiliation{Astronomy Unit, School of Physics and Astronomy, Queen Mary University of London, London E1 4NS, UK}
\email{t.haworth@qmul.ac.uk}

\author[0000-0002-5462-9387]{Rens Waters}
\affiliation{Department of Astrophysics/IMAPP, Radboud University, PO Box 9010, 6500 GL Nijmegen, The Netherlands}
\affiliation{SRON Netherlands Institute for Space Research, Niels Bohrweg 4, NL-2333 CA Leiden, the Netherlands}
\email{rens.waters@ru.nl}

\author[0000-0001-8068-0891]{Arjan Bik}
\affiliation{Department of Astronomy, Stockholm University\\
AlbaNova University Centre\\
106 91 Stockholm, Sweden}
\email{arjan.bik@astro.su.se}

\author[0000-0003-1413-1776]{Charles J. Law}
\affiliation{Minnesota Institute for Astrophysics, University of Minnesota, 116 Church St. SE, Minneapolis, MN 55455, USA}
\email{law06855@umn.edu}

\author[0000-0002-7501-9801]{Andrew J. Winter}
\affiliation{Astronomy Unit, School of Physics and Astronomy, Queen Mary University of London, London E1 4NS, UK}
\email{andrew.winter@qmul.ac.uk}

\author[0000-0003-4452-0588]{Vladimir S. Airapetian}
\affiliation{American University\\ 4400 Massachusetts Avenue NW\\ Washington, DC 20016 USA}
\affiliation{NASA/GSFC/SEEC\\ Greenbelt, MD, 20771, USA}
\email{vladimir.airapetian-1@nasa.gov}

\author[0000-0002-1566-389X]{Abygail R. Waggoner}
\affiliation{Department of Astronomy, University of Wisconsin-Madison, 475 N Charter St, Madison, WI 53706}
\email{awaggoner2@wisc.edu}

\author[0000-0002-0631-7514]{Michael A. Kuhn}
\affiliation{Centre for Astrophysics Research\\
Department of Physics, Astronomy and Mathematics\\
University of Hertfordshire\\
Hatfield, AL10 9UW, UK}
\email{m.kuhn@herts.ac.uk}

\author[0000-0003-2076-8001]{L. Ilsedore Cleeves}
\affiliation{University of Virginia, Charlottesville, VA 22904, USA}
\email{lic3f@virginia.edu}

\author[0000-0001-6010-6200]{Sergio A. Dzib}
\affiliation{Max-Planck-Institut fur Radioastronomie (MPIfR), Auf dem Hugel 69, 53121 Bonn, Germany}
\email{sdzib@mpifr.de}

\author[0000-0003-2303-0096]{Marco Padovani}
\affiliation{INAF--Osservatorio Astrofisico di Arcetri\\
Largo E. Fermi 5\\
50125 Firenze, Italy}
\email{marco.padovani@inaf.it}

\begin{abstract}
We present a systematic spectroscopic study of hydrogen Brackett-$\alpha$ (Br$\alpha$) and carbon monoxide (CO) fundamental-band (4.3-5.2 $\mu$m) emission from a homogeneous sample of low-mass (0.2-1.5 M$_{\odot}$) pre-main-sequence (PMS) stars in the Orion Nebula observed with JWST/NIRSpec-MOS as part of our Chandra-JWST project. Using 73 Br$\alpha$ and 46 CO spectra, we investigate correlations between warm-gas emission, stellar properties, accretion, and environment with censored correlation and regression analyses accounting for multiple testing. Br$\alpha$ luminosity correlates strongly with accretion luminosity ($L_{\rm acc}$), even after excluding sources within $2\arcmin$ of $\theta^1$~Ori~C, the Nebula's main ionizing O-type star, whose radiation may bias proplyd $L_{\rm acc}$ estimates. This supports Br$\alpha$ as a robust empirical tracer of accretion in PMS stars. Br$\alpha$ also correlates with visual extinction ($A_V$) and more weakly with stellar bolometric luminosity and mass. CO fundamental-band luminosities likewise correlate positively with $L_{\rm acc}$ and $A_V$, relationships that persist in a larger combined JWST/NIRSpec+MIRI sample spanning multiple star-forming regions. In contrast to Br$\alpha$, CO luminosity shows no significant dependence on basic stellar properties, suggesting that warm molecular gas is governed primarily by accretion and local disk conditions. The weaker correlations of the Spitzer/IRAC spectral slope with $L_{\rm acc}$ and $A_V$ suggest that warm inner-disk gas may evolve or disperse faster than optically thick dust. Overall, the CO results tentatively indicate declining warm-gas emission as disks evolve, modulated by accretion-powered thermal and ultraviolet excitation, although variations in emitting area inherited from natal core conditions may also contribute.
\end{abstract}



\section{Introduction} \label{sec_Introduction} 

Our coordinated Chandra and JWST/NIRSpec multi-object spectroscopy (MOS) observations of numerous disk-bearing pre-main-sequence (PMS) stars in the Orion Nebula were primarily designed to investigate the effects of stellar X-ray flares on protoplanetary disk chemistry \citep[hereafter Paper I;][]{Getman2026}. Beyond the flare-focused analysis of Paper I, these observations provide an opportunity to investigate the brightest spectral emission features in our dataset.

Specifically, we have measured Hydrogen Brackett-alpha, CO R-branch, and CO P-branch emission fluxes in JWST/NIRSpec spectra for a substantially larger sample of disk-bearing COUP-JWST stars\footnote{Throughout this paper, we designate sources as COUP-JWST,
where COUP refers to the X-ray source identification in the
Chandra Orion Ultradeep Project (COUP; \citealt{Getman2005}),
and JWST denotes the sequential source numbering adopted
for our JWST/NIRSpec program \#9292.} than was used in Paper I, including objects regardless of whether two-epoch spectra are available or X-ray flares are detected. Using this expanded sample, we investigate how Br-alpha and CO emission are related to one another and how they depend on stellar, accretion, disk, and environmental properties. Such an analysis may provide new insights into the physical origin, excitation mechanisms, and evolution of hydrogen and molecular CO emission in protoplanetary disks.

{\bf Hydrogen Brackett-alpha:} Although hydrogen recombination lines such as H$\alpha$, Pa$\beta$, and Br$\gamma$ are well-established empirical tracers of magnetospheric accretion in PMS stars, the physical origin and diagnostic potential of the 4.05 micron H I Br$\alpha$ transition remain comparatively poorly explored. Historically, ground-based observations of Br$\alpha$ have been severely limited by strong telluric absorption, restricting most studies to bright embedded protostars, Herbig Ae/Be stars, and massive young stellar objects. In these luminous systems, Br$\alpha$ has generally been interpreted as tracing ionized stellar winds, disk winds, compact H II regions, or shock-excited gas rather than magnetospheric accretion \citep[e.g.,][]{Persson1984,Bunn1992,Nisini1995,Drew1998}. More recently, JWST/MIRI observations have opened access to longer-wavelength H I recombination lines as potential diagnostics of accretion in young stellar objects \citep{Tofflemire2025,Shridharan2026,Reyes-Reyes2026}. Nevertheless, the applicability of such line-based accretion diagnostics can depend on stellar mass and evolutionary state, and whether Br$\alpha$ itself serves as a reliable empirical accretion indicator for low-mass (0.2--1.5 M$_{\odot}$) T Tauri stars remains largely untested. 

Early ground-based Br$\alpha$ measurements were obtained predominantly for luminous young stars, including the Herbig Ae/Be sample of \citet{Nisini1995}, who interpreted the infrared hydrogen emission lines using an ionized stellar-wind model. More recently, \citet{Komarova2020} used IRTF/SpeX observations of seven classical T Tauri stars to derive an initial empirical Br$\alpha$--$L_{\rm acc}$ relation, based on accretion luminosities estimated from Br$\gamma$ and archival UV-excess measurements. \citet{Testi2025} extended this approach using new IRTF/SpeX spectra of seven T Tauri stars and one Herbig Ae star in Taurus and Ophiuchus, supplemented by archival observations of 17 predominantly Herbig Ae/Be stars from \citet{Nisini1995}. For their combined Class II sample, they inferred $L_{\rm acc}$ from an established Br$\gamma$--$L_{\rm acc}$ calibration and obtained $\log(L_{\rm acc}/L_\odot)=(1.23\pm0.04)\log(L_{\rm Br\alpha}/L_\odot)+(4.41\pm0.11)$. 

Although these pioneering studies provide quantitative evidence that Br$\alpha$ can trace stellar accretion, their calibrations are based on small T Tauri samples as well as heterogeneous stellar populations and observing datasets. Our large, homogeneous JWST/NIRSpec survey of low-mass T Tauri stars in the Orion Nebula provides an independent assessment of Br$\alpha$ as an empirical accretion diagnostic. As shown below, our calibration of $L_{\rm acc}$ as a function of $L_{\rm Br\alpha}$ agrees closely with the relation derived by \citet{Testi2025}, despite differences in sample composition and statistical methodology.

\begin{figure*}[!t]
\centering
\includegraphics[width=0.85\textwidth]{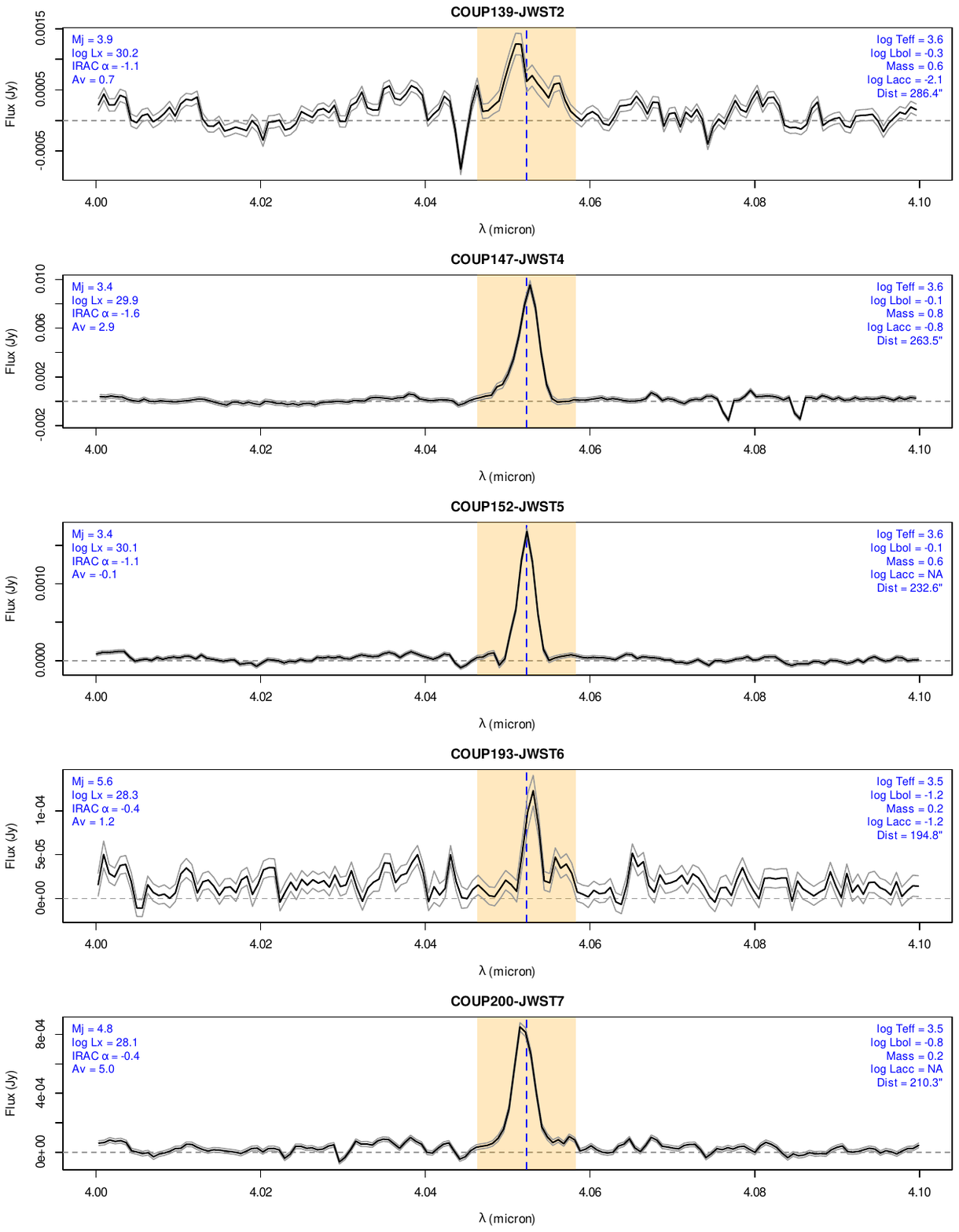}
\caption{Examples of the observed (i.e., prior to scaling to the Spitzer/IRAC flux level and correcting for extinction) JWST/NIRSpec spectra showing the H~I Br$\alpha$ line in COUP-JWST systems. The orange-shaded region marks the flux integration band, the vertical dashed line indicates the vacuum wavelength of Br$\alpha$, and the two gray curves show the $\pm1\sigma$ flux uncertainties. Panel annotations list the absolute $J$-band magnitude, X-ray luminosity, Spitzer/IRAC SED slope, visual extinction, stellar effective temperature, bolometric luminosity, and mass (Paper I); accretion luminosity \citep{Manara2012}; and projected angular distance from the main ionizing O-type star $\theta^1$~Ori~C. Only the first five sources are shown; a Figure Set containing all 73 COUP-JWST spectra included in the Br$\alpha$ analysis is available in the electronic edition. Spectra with non-detections of Br$\alpha$ emission are identified by red titles.} \label{fig_Br_alpha_Example}
\end{figure*}

\begin{figure*}[!t]
\centering
\includegraphics[width=0.85\textwidth]{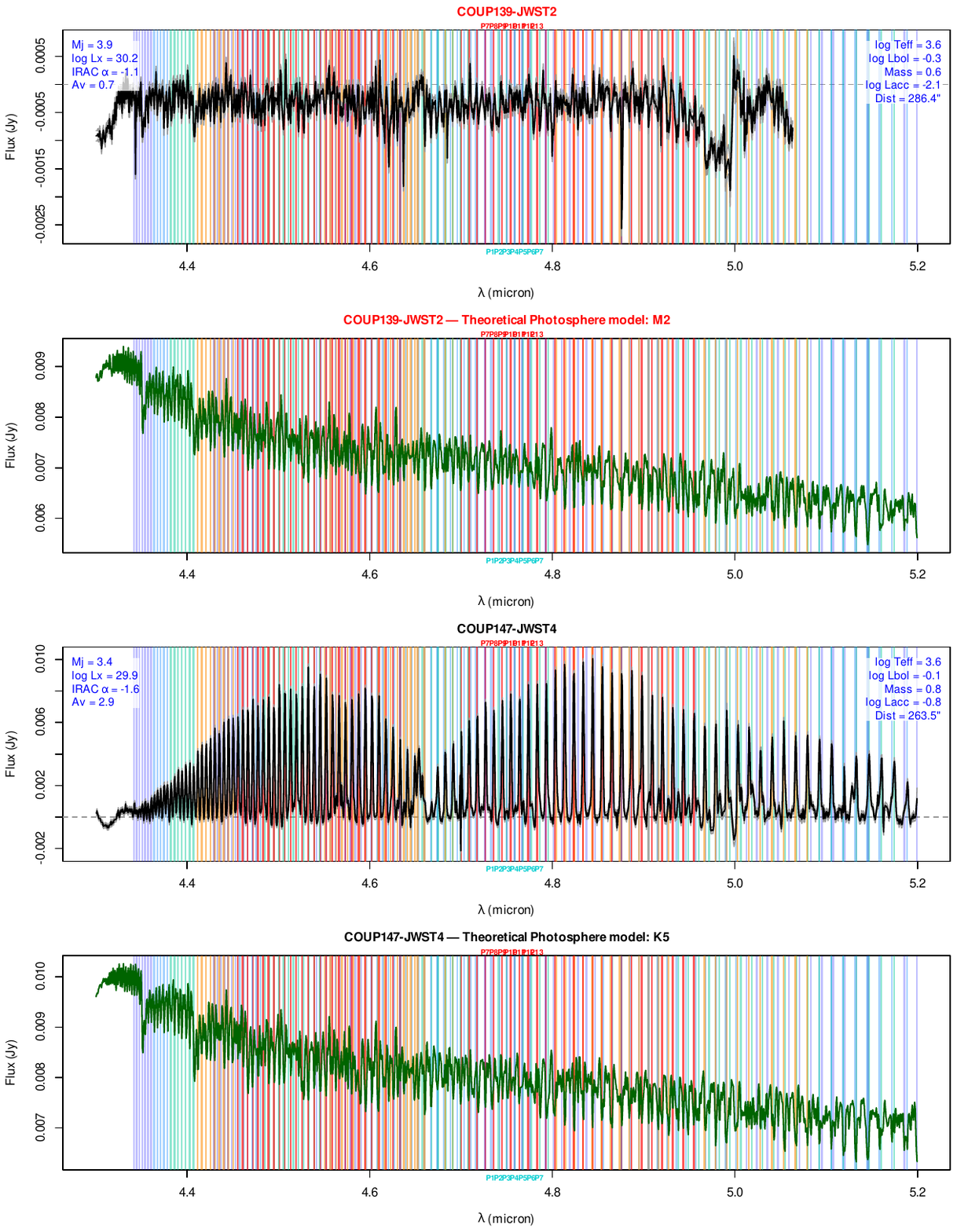}
\caption{Examples of the observed (i.e., prior to scaling to the Spitzer/IRAC flux level and correcting for extinction) JWST/NIRSpec spectra of COUP-JWST systems showing the CO spectral region (black curves). Corresponding theoretical stellar photosphere models (green curves) are from the BT-Settl CIFIST grid \citep{Allard2012}. Vertical lines mark theoretical CO transitions from the HITRAN database \citep{Gordon2022HITRAN2020}, color-coded according to line strength at $T_{\rm gas}=1200$ K \citep{Glassgold2004} and grouped into five quantiles from strongest to weakest (red, orange, teal, light blue, and blue). Figure legends are similar to those of Figure~\ref{fig_Br_alpha_Example}. For illustration, the red and teal tick marks label selected $^{12}$CO $v=1\rightarrow0$ fundamental ($P7$--$P13$) and $v=2\rightarrow1$ hot-band ($P1$--$P7$) transitions, respectively. Following \citet{Banzatti2015}, the fundamental line cores predominantly trace slow molecular disk winds, whereas the vibrationally excited hot-band transitions probe hot gas in the inner Keplerian disk. Only the first two sources are shown here; a Figure Set containing all 46 COUP-JWST NIRSpec spectra included in the CO analysis is available in the electronic edition of the paper. Spectra with CO emission non-detections are indicated by red figure titles.} \label{fig_CO_Example}
\end{figure*}

{\bf CO:} The fundamental ro-vibrational band ($\Delta v = 1$) of carbon monoxide CO spanning $4.3{-}5.2\ \mu\text{m}$ serves as a premier spectroscopic probe of the physical and chemical conditions within the inner several astronomical units of protoplanetary disks. High-resolution ground-based observations heavily focused on the $4.7\ \mu\text{m}$ CO fundamental band, establishing that the emission predominantly arises from warm (typically several hundred to ~1500 K), dense molecular gas within the terrestrial planet-forming zone. These early kinematic studies demonstrated that while the bulk of the line emission tracks Keplerian disk rotation, many profiles require an additional low-velocity component tracing slow molecular disk winds \citep{Najita2003,Blake2004,Salyk2011,Brown2013,Banzatti2015,Banzatti2022}.

Although CO is among the most readily detected molecular tracers in JWST spectra and remains the primary probe of warm molecular gas in the inner disk, recent demographic surveys have largely focused on chemical abundance diagnostics provided by species such as HCN, C$_{2}$H$_{2}$, CO$_{2}$, and H$_{2}$O \citep[e.g.,][]{vanDishoeck2023,Kamp2023,Henning2024,Arulanantham2025}. A comparable population-level analysis centered on CO luminosities and excitation has yet to be carried out. Space-based JWST/MIRI surveys are also limited to the P-branch wavelengths ($\gtrsim 4.9\ \mu\text{m}$). 

In this paper, we utilize a large, homogeneous sample of low-mass ($0.2{-}1.5\ M_{\odot}$) T Tauri stars in the Orion Nebula star-forming region observed with JWST/NIRSpec-MOS. While sacrificing the ultra-high spectral resolution of ground-based echelle instruments, this moderate-resolution survey provides the uniform sample size and complete wavelength coverage necessary to systematically evaluate the empirical scaling relations between the integrated CO fundamental-band luminosities and stellar and environmental properties. Such an analysis complements the pioneering VLT/CRIRES survey of \citet{Brown2013}, who found that vibrationally excited CO emission becomes increasingly prominent in systems with higher accretion luminosities, consistent with ultraviolet fluorescent excitation by the accretion shock. Their results provided some of the first evidence linking CO excitation to the accretion process. Here, we extend that framework by asking whether the luminosity of the CO fundamental band itself exhibits systematic relationships with stellar accretion and other stellar, disk, and environmental properties across a large, homogeneous population of Orion protoplanetary disks.

The current paper is organized as follows. Section~\ref{sec_methodology} summarizes the JWST/NIRSpec observations and data reduction described in Paper~I and presents the Br$\alpha$ and CO spectral measurements for the expanded Orion sample. Section~\ref{sec_results} examines correlations of Br$\alpha$ and CO luminosities with stellar, accretion, disk, and environmental properties, including a combined NIRSpec+MIRI analysis of CO emission across multiple star-forming regions. In Section~\ref{sec_discussion}, we discuss the physical origins and excitation mechanisms of Br$\alpha$ and CO emission and their implications for inner-disk gas evolution. Section~\ref{sec_conclusions} summarizes our principal findings. Supplementary catalogs of the Orion NIRSpec and comparison MIRI samples are provided in Appendix~\ref{sec_appendix_tables}, while Appendix~\ref{sec_appendix_multivariate_COUP_analysis} presents a multivariate correlation analysis of stellar and environmental properties in the larger Orion Nebula population.


\section{Methodology}\label{sec_methodology}

The James Webb Space Telescope (JWST; \citealt{Gardner2006,Rigby2023}) NIRSpec Multi-Object Spectroscopy (MOS; \citealt{Jakobsen2022}) observations of the Orion Nebula and the corresponding data reduction are described in Paper I. Briefly, the observations (program \#9292) employed the G395H/F290LP high-resolution grating/filter combination ($R \approx 2700$), covering 2.87--5.14~$\mu$m. Of the eight planned NIRSpec-MOS visits, three were successfully executed, yielding spectra for 84 disk-bearing PMS stars, including 44 observed at two epochs. Data reduction employed the official JWST calibration pipeline (version 1.20.2; \citealt{Bushouse2025}), with CRDS version 13.1.1 and calibration context \texttt{jwst\_1464.pmap}, together with a customized reduction procedure incorporating a cross-dispersion aperture correction for source-position-dependent slit losses. As discussed in Paper I, residual position-dependent pathloss effects produce systematic offsets in the absolute NIRSpec flux scale relative to Spitzer/IRAC photometry. For the integrated Br$\alpha$ and CO measurements analyzed here, we therefore use the IRAC photometry to place the NIRSpec spectra on an approximate absolute flux scale.

For the Br$\alpha$ and CO analyses, we adopted a single NIRSpec spectrum per source, selecting the epoch with the most complete wavelength coverage of both bands. When multiple non-truncated Micro-Shutter Array (MSA) spectra were available within that epoch, one was selected at random. Given the generally high signal-to-noise ratios of the Br$\alpha$ and CO features, combining multiple spectra was not necessary for the present population-level analysis.

Continuum-subtracted H I Br$\alpha$ fluxes were measured for 73 COUP-JWST stars (Figure~\ref{fig_Br_alpha_Example} and Table~\ref{tab_paperII_catalog} in Appendix~\ref{sec_appendix_tables}), including 58 detections and 15 non-detections. To mitigate residual JWST/NIRSpec pathloss and absolute flux-calibration uncertainties (Paper I), all line fluxes were scaled using a multiplicative correction factor ($CF$) defined as the ratio of the Spitzer/IRAC 4.5 $\mu$m flux density \citep{Megeath2012} to the median NIRSpec continuum flux density, measured with CTOOL \citep{Pontoppidan2026} over 4.4-4.6 $\mu$m. The fluxes were subsequently corrected for interstellar extinction using the \citet{Gordon2023} reddening law. Because no truly line-free local continuum exists near Br$\alpha$, upper limits for non-detections were conservatively adopted as $5\sigma$ of the flux uncertainty. Fluxes were converted to luminosities assuming a distance of 400 pc \citep{Getman2019}.

CO fundamental-band fluxes were measured for 46 COUP-JWST stars (Figure~\ref{fig_CO_Example} and Table~\ref{tab_paperII_catalog} in Appendix~\ref{sec_appendix_tables}), including 34 emission detections and 12 spectra dominated by CO absorption attributable to the stellar photosphere and/or weak absorption from the disks themselves. After continuum subtraction, fluxes were integrated over three NIRSpec wavelength intervals (4.420--4.477, 4.477--4.6442, and 4.678--5.000 $\mu$m), with the number of usable spectra decreasing toward longer wavelengths because of NIRSpec-MOS wavelength truncation. The gap between 4.6442 and 4.678 $\mu$m excludes the central region between the CO $R$- and $P$-branches near 4.67 $\mu$m and avoids contamination from the H~I Pfund-$\beta$ ($7\rightarrow5$) line at 4.6538 $\mu$m \citep{Salyk2013}. For comparison with JWST/MIRI observations, an additional integration over the 4.90--5.11 $\mu$m interval was performed. For spectra dominated by CO absorption, the continuum was estimated using a rolling 90th-percentile filter implemented with the zoo package in R \citep{Zeileis2005}, rather than CTOOL, providing a more appropriate representation of the stellar continuum. As for Br$\alpha$, the measured CO fluxes were corrected to the IRAC 4.5 $\mu$m absolute flux scale, dereddened using \citet{Gordon2023}, converted to luminosities assuming $d=400$ pc, and assigned $5\sigma$ upper limits in cases of non-detection. The 12 absorption-dominated spectra are treated as upper limits on the CO emission in the subsequent statistical analyses.

\section{Results} \label{sec_results}

\subsection{NIRSpec HI Br$\alpha$} \label{subsec_results_Bralpha}

Prior to examining the Br$\alpha$ correlations, the reader is referred to Appendix~\ref{sec_appendix_multivariate_COUP_analysis}, which summarizes the covariance structure of stellar and environmental properties in the much larger parent Chandra Orion Ultradeep Project \citep[COUP;][]{Getman2005} sample. That analysis shows that quantities associated with circumstellar evolution, including X-ray median photon energy, visual extinction, IRAC spectral slope, and accretion luminosity, are moderately correlated with one another, while the fundamental stellar parameters (stellar mass, effective temperature, bolometric luminosity, and X-ray luminosity) likewise form a strongly interrelated group. Aside from their mutual correlation with accretion luminosity, however, these two sets of parameters exhibit only weak coupling, providing a useful framework for interpreting the Br$\alpha$ correlations presented below.

To evaluate relationships between Br$\alpha$ luminosity and stellar or environmental properties, we performed a series of eight bivariate censored Kendall's $\tau$ correlation analyses using the generalized Kendall's $\tau$ statistic for censored data and estimated the corresponding regression slopes using the Akritas--Theil--Sen (ATS) estimator, as implemented in the `cenken` function of the NADA2 package for R \citep{Helsel2011,NADA2_package}. Br$\alpha$ non-detections were incorporated as $5\sigma$ upper limits. Because eight correlation tests were performed, the resulting $p$-values were adjusted for multiple comparisons using the \citet{BenjaminiHochberg1995} False Discovery Rate (FDR) procedure \citep[see also][]{Miller2001}. Several of the tested stellar parameters (e.g., X-ray luminosity, stellar mass, and $L_{\rm bol}$) are physically and empirically correlated along pre-main-sequence evolutionary tracks and therefore do not represent independent astrophysical hypotheses. Because the effective number of independent astrophysical variables is smaller than the number of statistical tests, the FDR correction provides a conservative assessment of these related stellar trends. We distinguish between highly significant correlations ($q_{\rm FDR}<0.01$) and suggestive correlations ($0.01\leq q_{\rm FDR}<0.05$), interpreting the latter more cautiously.

\begin{figure*}[!t]
\centering
\includegraphics[width=0.90\textwidth]{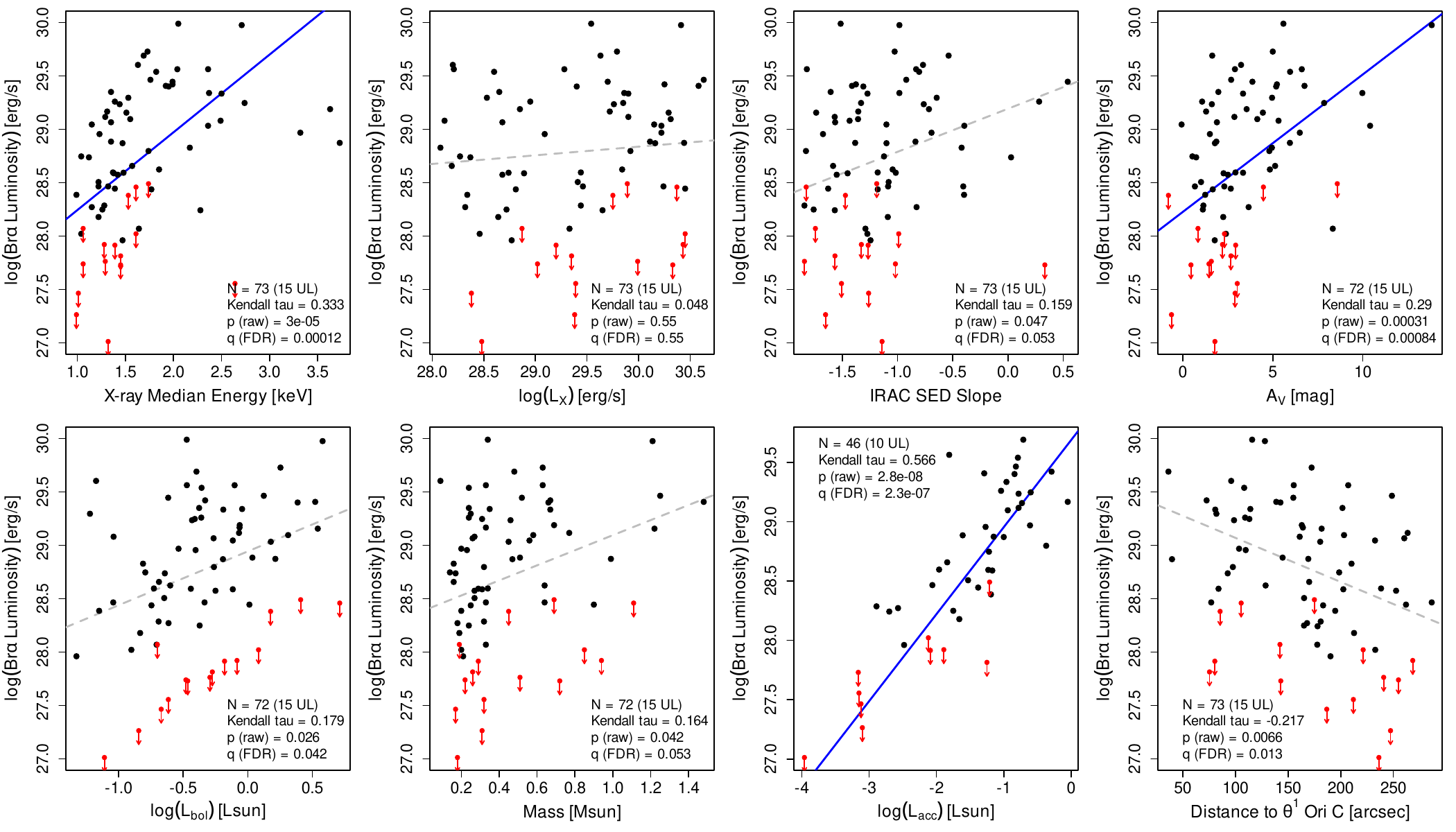}
\caption{Pairwise comparison plots of H I Br$\alpha$ luminosity versus stellar and environmental properties: X-ray median photon energy, X-ray luminosity, Spitzer/IRAC spectral slope, visual extinction, stellar bolometric luminosity, stellar mass, accretion luminosity, and projected distance from $\theta^1$~Ori~C. Each panel reports the total number of COUP-JWST systems analyzed (with the number of upper limits in parentheses) together with the results of the censored Kendall's $\tau$ correlation analysis, including false discovery rate-adjusted $p$-values ($q_{\rm FDR}$). Upper limits are indicated by red downward arrows. Akritas--Theil--Sen (ATS) regression lines are shown as solid blue for highly statistically significant correlations ($q_{\rm FDR} < 0.01$) and as dashed gray otherwise.} \label{fig_Br_alpha_analysis1}
\end{figure*}

\begin{figure*}[!t]
\centering
\includegraphics[width=0.90\textwidth]{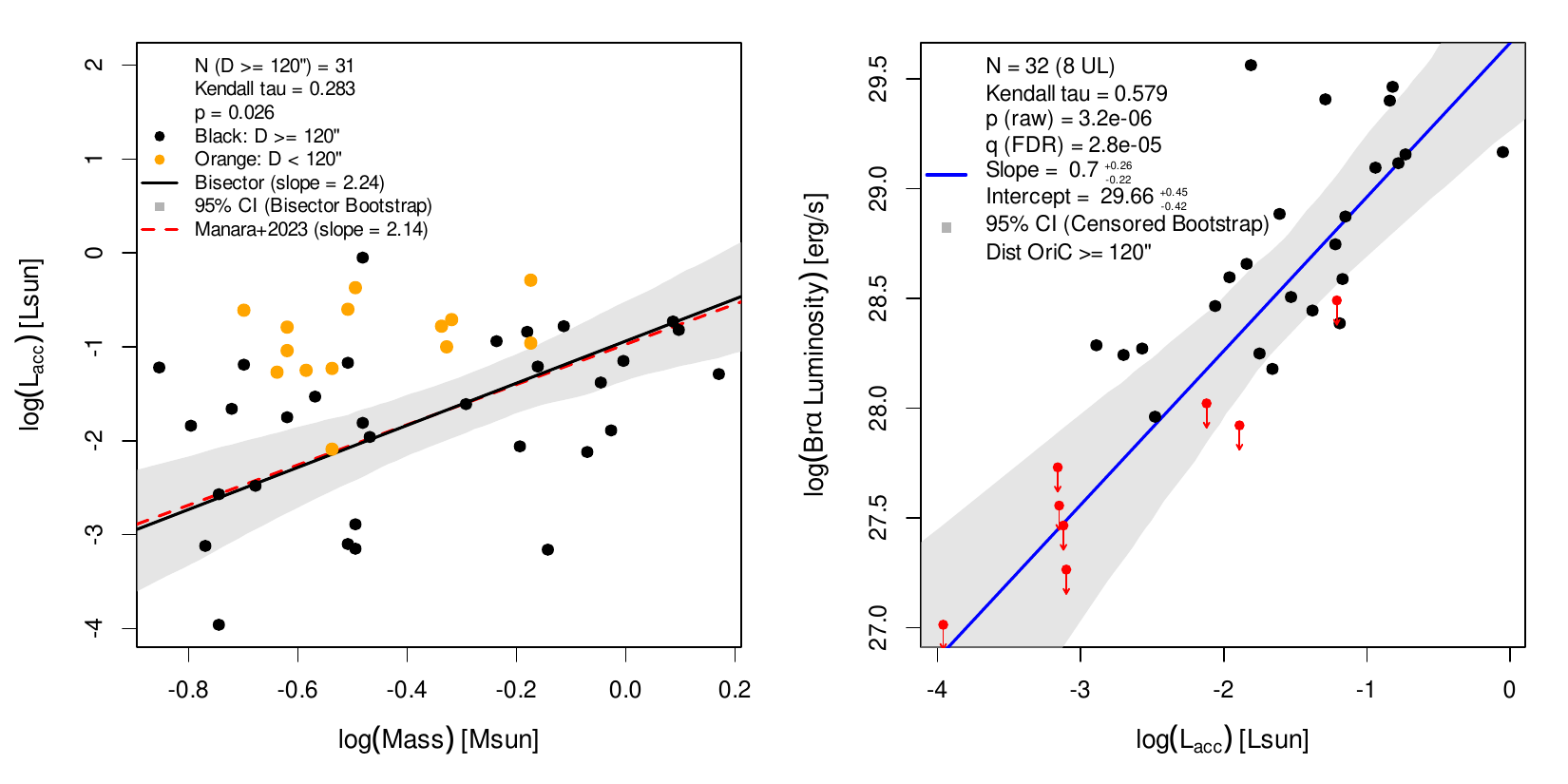}
\caption{(Left panel): Accretion luminosity as a function of stellar mass for the COUP-JWST sample analyzed in this study. Stars with projected distances $R<2\arcmin$ from $\theta^1$~Ori~C are shown as orange points, while those with $R\ge 2\arcmin$ are shown as black points. The solid black line and gray shaded region denote the OLS bisector regression fit \citep{Isobe1990} and its confidence interval for the $R \ge 2\arcmin$ subsample. For comparison, the dashed red line shows the OLS bisector regression derived in this work from the compilation of PMS stars in nearby low-mass star-forming regions presented by \citet{Manara2023}. The panel legends report the results of the Kendall's $\tau$ rank correlation analysis for the $R \ge 2\arcmin$ subsample together with the slopes of the two regression relations. (Right panel): Updated version of the Br$\alpha$ luminosity versus accretion luminosity relation from Figure~\ref{fig_Br_alpha_analysis1} after excluding stars within $2\arcmin$ of $\theta^1$~Ori~C. The panel legend reports the results of the censored Kendall's $\tau$ correlation analysis together with the slope and intercept of the Akritas--Theil--Sen regression (solid blue line).} \label{fig_Br_alpha_analysis2}
\end{figure*}

The Br$\alpha$ luminosities exhibit a highly statistically significant positive correlation with stellar accretion luminosity ($\tau=0.57$, $q_{\rm FDR}=2.3\times10^{-7}$; Figure~\ref{fig_Br_alpha_analysis1}). The accretion luminosities adopted here are taken from \citet{Manara2012}, who derived $L_{\rm acc}$ from HST U-band excess and H$\alpha$ measurements. Although this catalog provides by far the largest homogeneous set of accretion luminosities available for the Orion Nebula, our multivariate analysis of the much larger parent COUP sample (Appendix~\ref{sec_appendix_multivariate_COUP_analysis}) indicates that these $L_{\rm acc}$ estimates exhibit a stronger dependence on projected distance from $\theta^1$~Ori~C than any other stellar or environmental parameter considered. This behavior raises the possibility of a systematic observational bias affecting stars in the central Trapezium region.

A plausible explanation is contamination of the U-band excess and H$\alpha$ measurements by compact photoionized proplyd emission. While the diffuse Orion Nebula background can largely be removed owing to the angular resolution of the HST imaging, the bright ionized cusps surrounding many proplyds are only marginally separable from the stellar point-spread function and emit strongly in both the Balmer continuum and H$\alpha$. Such emission could artificially increase the inferred accretion luminosities of stars projected close to $\theta^1$~Ori~C.

This possibility is examined in Figure~\ref{fig_Br_alpha_analysis2} (left), which compares the Orion Nebula COUP-JWST $L_{\rm acc}$--$M_\star$ relation with the independent compilation of \citet{Manara2023} for nearby low-mass star-forming regions lacking O-type stars (e.g., $\rho$ Oph, Chamaeleon, Corona Australis, Lupus, and Taurus). Stars located within $2$ arc-minutes of $\theta^1$~Ori~C (orange symbols) are systematically displaced toward higher accretion luminosities relative to the \citet{Manara2023} relation (red dashed line). After excluding these central objects, the OLS bisector fit to the remaining Orion Nebula stars (black line) becomes consistent with the reference relation.

The choice of $R=2\arcmin$ as the boundary of the central region is motivated empirically by the radial behavior of the variable-line population identified in Paper~I. In particular, the number of highly variable narrow NIRSpec emission features at wavelengths below 4.3~$\mu$m shows a strong negative correlation with projected distance from $\theta^1$~Ori~C, with the decline becoming markedly steeper at $R<2\arcmin$. This change in slope suggests that the innermost $\sim2\arcmin$ represents a region of enhanced environmental influence from $\theta^1$~Ori~C, potentially associated with the higher incidence of externally irradiated proplyd structures. We therefore adopt $R=2\arcmin$ as an empirical boundary for the central region rather than as a sharp physical boundary of the nebula.

Figure~\ref{fig_Br_alpha_analysis2} (right) presents the Br$\alpha$--$L_{\rm acc}$ correlation after excluding the central ($R<2\arcmin$) Orion Nebula stars. The correlation remains highly statistically significant ($\tau=0.58$, $q_{\rm FDR}=2.8\times10^{-5}$), demonstrating that the association between Br$\alpha$ luminosity and accretion luminosity is not driven by the potentially biased central sources. Among all stellar and environmental parameters considered, accretion luminosity therefore remains the strongest predictor of Br$\alpha$ emission. The corresponding Akritas--Theil--Sen (ATS) regression is

$\log(L_{\rm Br\alpha}) = 29.66^{+0.45}_{-0.42} + (0.70^{+0.26}_{-0.22}) \times \log(L_{\rm acc}/L_{\odot}),$

where $L_{\rm Br\alpha}$ is expressed in erg s$^{-1}$ and the asymmetric uncertainties represent 95\% confidence intervals estimated using a non-parametric censored bootstrap procedure.

To facilitate comparisons with published hydrogen-line accretion diagnostics (e.g., \citealt{MarleauAoyama2022, Testi2025,Shridharan2026}), we also performed an independent censored ATS regression with the dependent and independent variables exchanged, retaining the Br$\alpha$ upper limits as censored values of the independent variable. For the same 32 sources, we obtain

$\log(L_{\rm acc}/L_{\odot}) = 4.18^{+1.65}_{-1.82} + (1.19^{+0.29}_{-0.37}) \times \log(L_{\rm Br\alpha}/L_{\odot}),$

where the uncertainties again represent 95\% bootstrap confidence intervals. We emphasize that this relation is obtained from a separate censored regression and {\it is not an algebraic inversion} of the original ATS relation.

We also find a highly statistically significant, moderate positive correlations with X-ray median energy (a surrogate for gas absorption) and visual extinction ($\tau \sim 0.30$, $q_{\rm FDR} < 0.001$). Notably, a marginally significant positive correlation is already present between $A_V$ and the observed (i.e., prior to extinction correction) Br$\alpha$ luminosities, indicating that the relationship is not introduced by the dereddening procedure.

As a check that the empirical IRAC/NIRSpec flux-recovery correction does not introduce systematic trends into the subsequent correlation analysis, we tested the Br$\alpha$ correction factor ($CF$) against the eight stellar and environmental properties considered above. We find no statistically significant correlations after accounting for multiple testing ($q_{\rm FDR}>0.05$ in all cases).

Beyond the primary correlations of Br$\alpha$ with $L_{\rm acc}$ and $A_V$, we identify a weaker, suggestive positive correlation with stellar bolometric luminosity ($\tau=0.18$, $q_{\rm FDR}=0.042$). Stellar mass shows a similar but slightly weaker positive trend ($\tau=0.16$, $q_{\rm FDR}=0.053$), which does not meet our adopted significance threshold. Because these stellar quantities are themselves strongly correlated along PMS evolutionary tracks, these trends may reflect a common underlying dependence of Br$\alpha$ luminosity on stellar evolution rather than independent relationships. The Spitzer/IRAC spectral slope also shows a weak positive trend ($\tau=0.16$, $q_{\rm FDR}=0.053$), although it likewise falls just above the significance threshold.

\begin{figure*}[!t]
\centering
\includegraphics[width=0.90\textwidth]{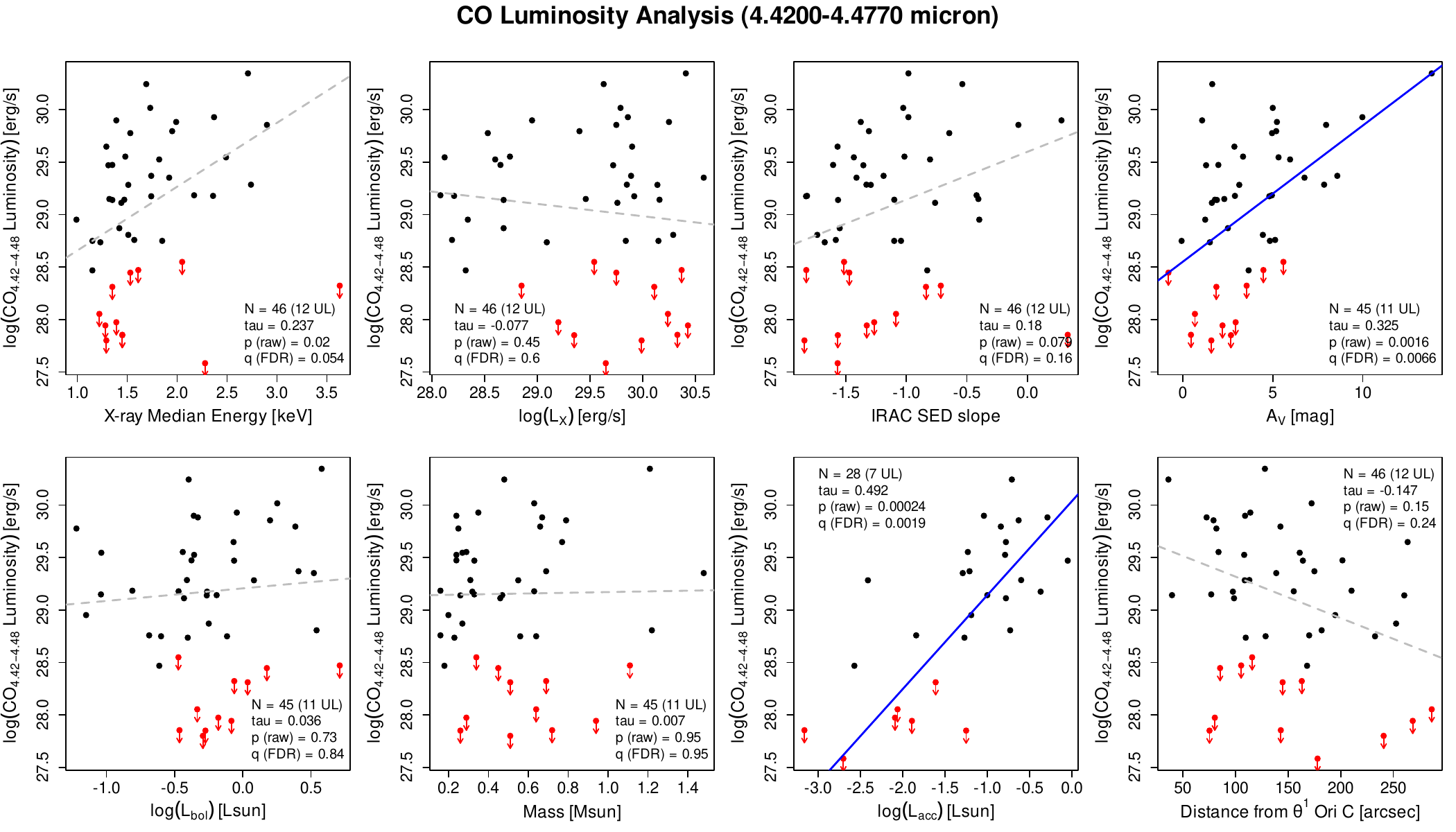}
\caption{Pairwise comparison plots of the CO luminosity versus various stellar and environmental properties including X-ray median photon energy, X-ray luminosity, Spitzer/IRAC spectral slope, visual extinction, bolometric luminosity, stellar mass, accretion luminosity, and projected distance from $\theta^1$~Ori~C. The legends report the results of the censored Kendall's $\tau$ correlation analysis. A Figure Set containing results for other CO bands is available in the electronic edition of the paper.} \label{fig_CO_analysis1}
\end{figure*}

No statistically significant relationship is found between Br$\alpha$ luminosity and stellar X-ray luminosity ($\tau=0.05$, $q_{\rm FDR}=0.55$). Finally, Br$\alpha$ luminosity exhibits a weak but suggestive anti-correlation with projected distance from $\theta^1$~Ori~C ($\tau=-0.22$, $q_{\rm FDR}=0.013$), indicating that Br$\alpha$ emitters tend to be somewhat more luminous in the central regions of the cluster. We note, however, that this weak radial trend could also be influenced by contamination from externally photoionized proplyds in the central Trapezium. If compact ionization fronts contribute Br$\alpha$ emission within the NIRSpec aperture, they would preferentially enhance the measured Br$\alpha$ luminosities of stars projected close to $\theta^1$~Ori~C.

\subsection{NIRSpec CO} \label{subsec_results_CO}

The principal result is that CO fundamental-band luminosities exhibit statistically significant positive correlations with stellar accretion luminosity. In the full sample, the two shorter-wavelength CO bands, 4.42--4.477 and 4.477--4.644~$\mu$m, show highly significant correlations after accounting for the eight simultaneous tests ($q_{\rm FDR}<0.01$), with moderate-to-strong generalized Kendall correlations ($\tau \sim 0.5$; Figure~\ref{fig_CO_analysis1}). These results suggest that the strength of the CO fundamental emission is linked to the accretion-powered activity of the central PMS stars.

\begin{figure*}[!t]
\centering
\includegraphics[width=0.90\textwidth]{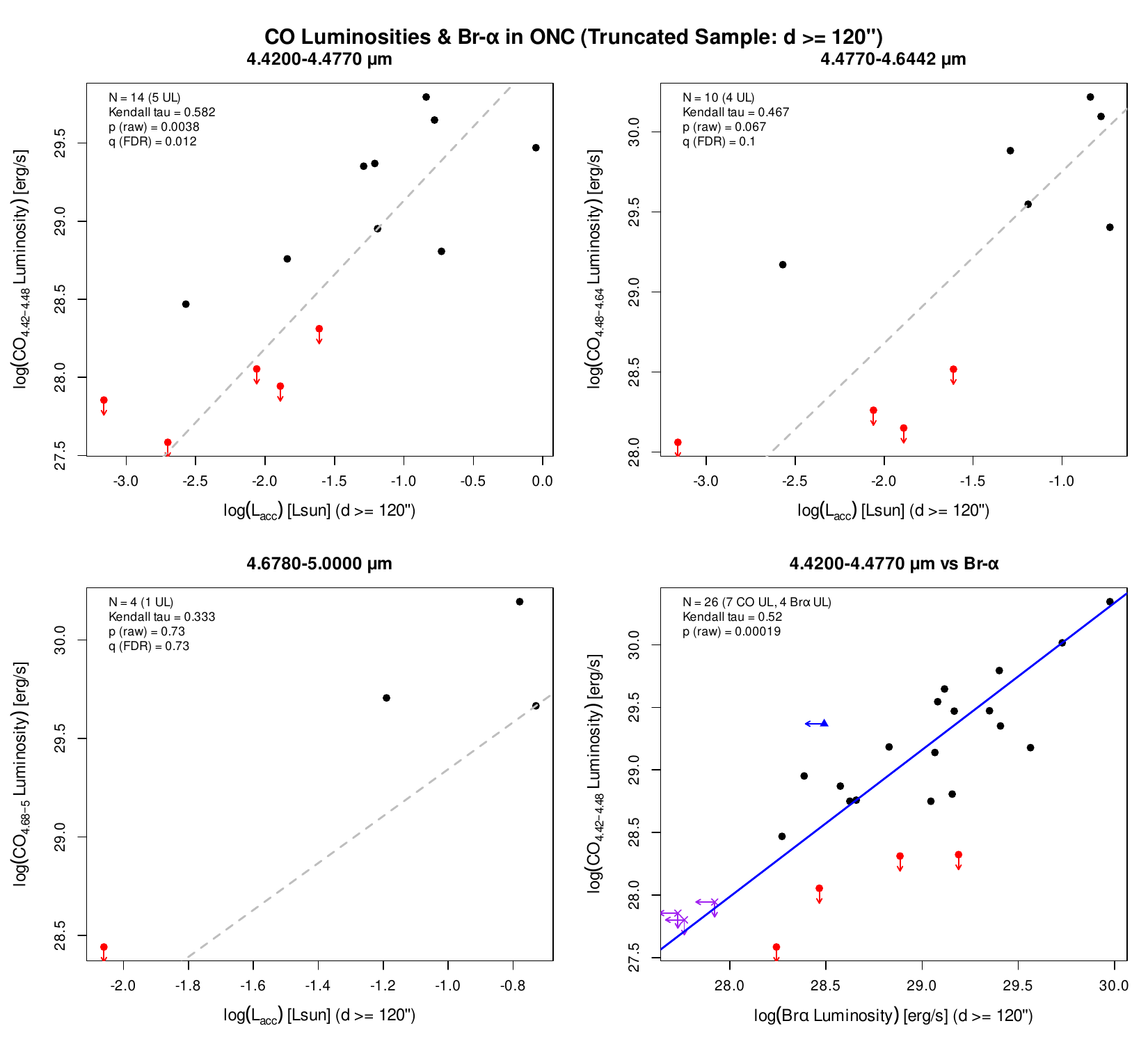}
\caption{The first three panels show CO luminosity versus stellar accretion luminosity for the three independent CO bands, using the same symbols, lines, and legend conventions as in Figure~\ref{fig_CO_analysis1}, but restricting the sample to stars located at projected distances $R>2\arcmin$ from $\theta^1$~Ori~C. The fourth panel shows the luminosity of the 4.42-4.477~$\mu$m CO band, which has the largest number of CO measurements, versus Br$\alpha$ luminosity for the same spatially restricted sample. Purple crosses mark stars with upper limits on both CO and Br$\alpha$ luminosities; blue triangles and red points mark stars with upper limits only on Br$\alpha$ and CO luminosities, respectively. The fourth-panel legend reports the raw p-value but not the FDR-corrected $q$-value because this panel tests a separate hypothesis involving a single pair of variables and therefore does not form part of the multiple-testing set used for the first three panels.} \label{fig_CO_analysis2}
\end{figure*}

However, as discussed in the preceding section, the $L_{\rm acc}$ estimates may be systematically affected for stars within $\sim 2\arcmin$ of $\theta^1$~Ori~C, where proplyd ionization fronts and spatially variable nebular emission can contaminate the photometry used to infer accretion luminosities. We therefore repeated the analysis after restricting the sample to $R>2\arcmin$ from $\theta^1$~Ori~C (Figure~\ref{fig_CO_analysis2}). The correlation remains particularly clear in the 4.42--4.477~$\mu$m band, which provides the largest CO sample: $\tau=0.6$ with $q_{\rm FDR}=0.012$. Although just above our adopted $q_{\rm FDR}<0.01$ threshold for a highly significant correlation, this result is statistically suggestive and substantially stronger in effect size than expected from a random association. The corresponding trends in the 4.477--4.644 and 4.678--5.000~$\mu$m bands are weaker ($\tau \sim 0.3$--0.5) and are not statistically significant given their smaller samples ($q_{\rm FDR}>0.05$). Thus, the evidence for a CO--accretion connection persists after removing the innermost, potentially contaminated region, but is currently driven primarily by the best-sampled CO band.

CO luminosity also correlates positively with visual extinction. Both shorter-wavelength CO bands show highly significant correlations ($q_{\rm FDR}<0.01$) with $A_V$, with moderate Kendall coefficients ($\tau \sim 0.35$). The recurrence of this relation across two independent spectral intervals argues against it being a statistical peculiarity of a single CO band. 

A positive relation is also found between CO luminosity and the COUP X-ray median photon energy ($ME$), a quantity sensitive to both intervening absorption and the characteristic temperature of the stellar coronal plasma. For the intermediate 4.477--4.644~$\mu$m band, the correlation is statistically significant ($\tau=0.30$), $q_{\rm FDR}=0.0094$; $N=37$). In contrast, the larger 4.42--4.477~$\mu$m sample ($N=46$) yields a weaker, statistically non-significant relation ($\tau=0.237$, $q_{\rm FDR}=0.054$). This difference is noteworthy because $ME$ is not a pure measure of absorption: it can also increase during strong X-ray flares as the coronal plasma becomes hotter. Thus, temporal variability in the X-ray emission may add scatter to, or partially obscure, an underlying relation between CO emission and the absorbing gas column. The $ME$ results therefore provide at most tentative evidence for an environmental/absorption connection.

Equally notable is what is not observed. In contrast to Br$\alpha$, for which the preceding analysis finds weak suggestive trends with stellar luminosity and mass, none of the CO bands shows a statistically significant dependence on stellar mass, bolometric luminosity, or stellar X-ray luminosity. The absence of these correlations, together with the positive CO--$L_{\rm acc}$ and CO--$A_V$ relations, indicates that the observed CO fundamental emission is not primarily controlled by global stellar properties. Instead, its strength appears to be more closely connected to accretion activity and to the local circumstellar environment, consistent with CO fundamental emission tracing the physical conditions of warm molecular gas in the inner disk.

\begin{figure*}[!t]
\centering
\includegraphics[width=0.90\textwidth]{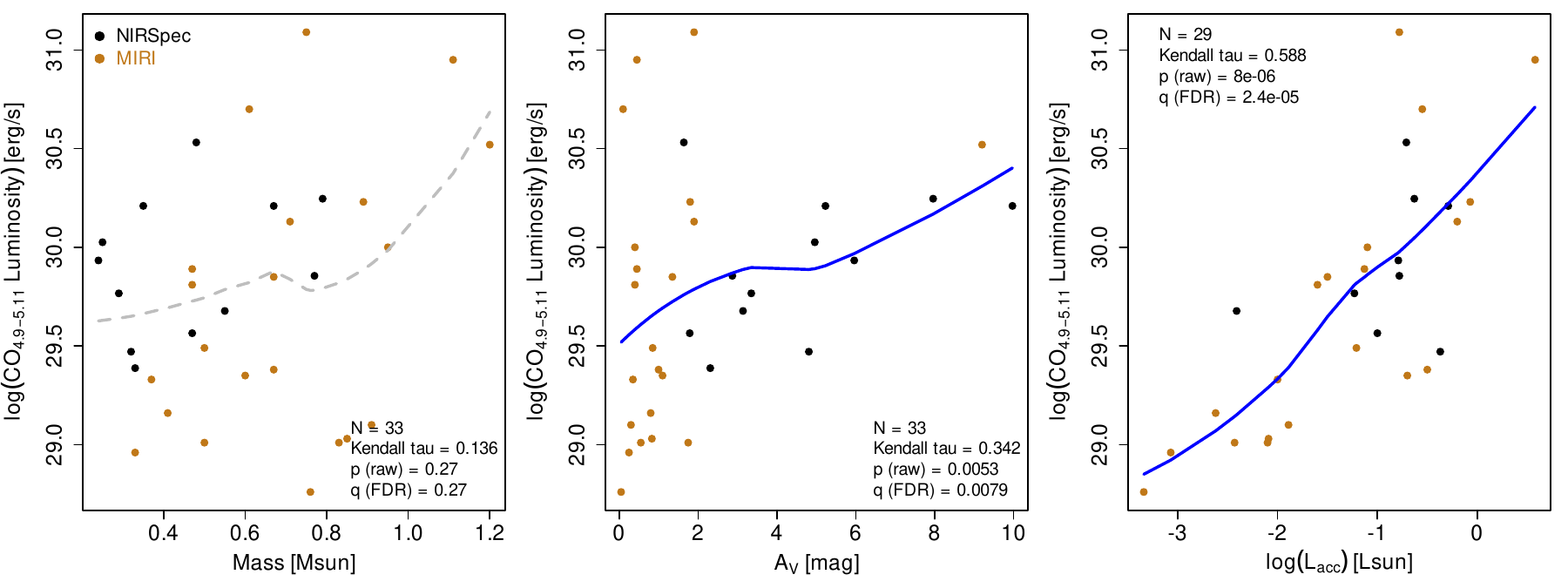}
\caption{Pairwise comparison plots of the CO fundamental-band luminosity integrated over the common 4.90--5.11 $\mu$m wavelength interval versus stellar mass, visual extinction, and accretion luminosity for the combined sample of JWST/NIRSpec (black symbols) and JWST/MIRI (orange symbols) PMS stars. The legends report Kendall's $\tau$ correlation statistics, and the curves show locally estimated scatterplot smoothing (LOESS) fits, which trace trends through locally weighted regressions without assuming a global functional form \citep{Cleveland1979}.} \label{fig_CO_analysis3}
\end{figure*}

The last panel of Figure~\ref{fig_CO_analysis2} shows that, even after restricting the COUP--JWST sample to systems located at $R>2\arcmin$ from $\theta^1$~Ori~C, the CO luminosity integrated over the 4.42--4.48~$\mu$m band remains strongly correlated with Br$\alpha$ luminosity $\tau=0.50$, $p_{\rm raw}=1.9\times10^{-4}$. Because the two quantities are measured from the same JWST/NIRSpec spectra at the same observing epochs, this correlation is not subject to the additional scatter that could arise from combining measurements obtained at different epochs, instruments, observing or methodological conditions. The most likely origin of the correlation is nevertheless not a direct causal connection between the CO and Br$\alpha$ emitting regions. Rather, both luminosities independently increase with stellar accretion luminosity and with source extinction, suggesting that their mutual correlation may arise because both trace common underlying properties of the disk-bearing systems. In particular, the common dependence on accretion activity points to a connection between the strength of the warm molecular CO emission and the processes that power hydrogen recombination-line emission, while the shared dependence on extinction may reflect systematic changes in the circumstellar environment and evolutionary state of the disks.

As a final robustness check, we assessed the potential influence of unresolved stellar multiplicity on the correlations presented in Figures 3 and 5, which include the full Br$\alpha$ and CO samples, respectively. Among the 75 unique sources in these samples, 71 have valid Gaia DR3 renormalized unit weight error (RUWE) measurements, of which four (COUP 152, 205, 1324, and 1342) exhibit RUWE $\geq 1.4$, potentially indicating unresolved companions, although source crowding, nebulosity, and other astrometric systematics can also produce elevated RUWE values. In addition, \citet{Kounkel2016} identify one further source (COUP 1193) as radial-velocity variable, while none of the binaries identified by \citet{Duchene2018} overlap with our sample. Excluding these five multiplicity-flagged sources produces only minor changes in the principal correlations: Kendall's $\tau$ for Br$\alpha$ versus $L_{\rm acc}$ changes from 0.566 to 0.560, and for the three CO bands from 0.412--0.492 to 0.412--0.481. Similarly, the Br$\alpha$--$A_V$ correlation changes from $\tau=0.290$ to 0.314, while the CO--$A_V$ correlations remain comparable ($\tau=0.325$--0.382 versus 0.334--0.407). The absence of significant correlations with X-ray luminosity is also unaffected. Additional undetected binaries could contribute blended Br$\alpha$ and CO emission and bias the inferred stellar masses, bolometric and accretion luminosities, and X-ray luminosities through unresolved contributions from multiple components, potentially introducing additional scatter or systematic offsets in these relationships. Nevertheless, the stability of the correlations after excluding the identified multiplicity candidates suggests that these objects do not drive our principal findings.



\subsection{NIRSpec+MIRI CO} \label{subsection_results_nirspec_miri}

To extend the NIRSpec CO analysis beyond Orion, we assembled a comparison sample of 21 disk-bearing PMS stars observed with JWST/MIRI. Twenty reside in nearby star-forming regions (Taurus, Lupus, Chamaeleon~I, and Upper Centaurus--Lupus), while XUE~1 belongs to the more distant massive star-forming region NGC~6357. Their MIRI spectra were obtained through the MINDS \citep{Henning2024}, JDISCS \citep{Pontoppidan2024,Arulanantham2025}, and XUE \citep{Ramirez-Tannus2023,Ramirez-Tannus2025} programs. Reduced and calibrated spectra were provided by collaborators from these programs, following the procedures described by \citet{Henning2024} (MINDS), \citet{Pontoppidan2024} (JDISCS), and \citet{Ramirez-Tannus2023,Ramirez-Tannus2025} (XUE). CO fluxes were measured by integrating the continuum-subtracted spectra over 4.90--5.11~$\mu$m (Table~\ref{tab_paperII_miri} in Appendix~\ref{sec_appendix_tables} and Frediani et al., in prep.).

Combining these 21 MIRI targets with 12 COUP--JWST sources having complete NIRSpec P-branch coverage and no contamination from stellar absorption yields a 33-source sample. The combined analysis largely reinforces the NIRSpec-only results. The strongest correlation is again between CO fundamental-band luminosity and stellar accretion luminosity, with a highly significant Kendall coefficient of $\tau=0.59$ ($q=2.4\times10^{-5}$), identifying accretion luminosity as the strongest correlate of CO emission (Figure~\ref{fig_CO_analysis3}). A significant positive correlation is also found with visual extinction ($\tau=0.34$, $q=0.008$), consistent with the Orion trends, whereas no statistically significant correlation with stellar mass is detected.

These results suggest that accretion-related heating and ultraviolet irradiation play important roles in exciting the CO fundamental emission. The combined NIRSpec+MIRI analysis thus supports a picture in which the 4.3--5.2~$\mu$m CO emission is primarily linked to inner-disk gas conditions and accretion activity, with stellar mass showing no detectable independent association in the present sample.

\section{Discussion} \label{sec_discussion}

\subsection{H I Br$\alpha$} \label{subsec_discussion_Bralpha}

The dominant result of this analysis is the remarkably strong correlation between the Br$\alpha$ luminosity and stellar accretion luminosity. Although the $L_{\rm acc}$ values adopted here may be systematically elevated for stars close to $\theta^1$~Ori~C because of contamination by compact photoionized proplyd emission, the Br$\alpha$--$L_{\rm acc}$ correlation remains highly significant after excluding these central sources. This robustness argues that the observed relationship is not produced by the same environmental bias that affects a subset of the accretion measurements. This finding places Br$\alpha$ alongside H$\alpha$, Pa$\beta$, Br$\gamma$, and other hydrogen recombination lines as a robust empirical tracer of mass accretion in PMS stars. The observed relationship is consistent with an origin of the Br$\alpha$ emission in dense hydrogen gas associated with magnetospheric accretion flows, where gas is ionized and heated as it is channeled from the inner disk onto the stellar surface. Although the precise formation region of the Br$\alpha$ line has remained uncertain since the pioneering work of \citet{Persson1984}, our results strongly support the interpretation that its luminosity in PMS stars is governed primarily by the accretion process.

Our independently derived censored ATS regression of $L_{\rm acc}$ versus $L_{\rm Br\alpha}$ is in remarkably good agreement with the empirical calibration reported by \citet{Testi2025}. In particular, our fitted slope of $1.19^{+0.29}_{-0.37}$ and intercept of $4.18^{+1.65}_{-1.82}$, with all luminosities expressed in solar units and uncertainties representing 95\% bootstrap confidence intervals, are consistent with the corresponding values of $1.23\pm0.04$ and $4.41\pm0.11$ reported by \citet{Testi2025}, who estimated their coefficient uncertainties by propagating measurement errors and uncertainties in the adopted Br$\gamma$--accretion luminosity calibration. The two relations predict nearly identical accretion luminosities over the range $\log(L_{\rm Br\alpha}/L_\odot)\sim-5$ to $-3$, differing by less than approximately 0.1 dex. This agreement is particularly noteworthy given the substantial differences in stellar populations and analysis methods. The \citet{Testi2025} calibration combines a small number of T Tauri stars with predominantly higher-mass Herbig Ae/Be objects and derives $L_{\rm acc}$ indirectly from Br$\gamma$ luminosities, whereas our regression is based on a relatively homogeneous sample of low-mass PMS stars and explicitly accounts for Br$\alpha$ non-detections through censored statistics. The close agreement between these independent calibrations strengthens the evidence that Br$\alpha$ luminosity provides a useful empirical tracer of stellar accretion across different star-forming environments and stellar populations.

Our results are also consistent with recent JWST studies demonstrating that longer-wavelength hydrogen recombination lines provide reliable diagnostics of stellar accretion. In particular, \citet{Tofflemire2025,Shridharan2026} showed that mid-infrared H I transitions correlate closely with independently measured accretion properties across diverse samples of young stellar objects. The present analysis extends this picture to the 4.05 $\mu$m Br$\alpha$ transition, providing evidence from a large, homogeneous JWST/NIRSpec survey of Orion disks that Br$\alpha$ likewise serves as an excellent empirical accretion indicator.

An important advantage of Br$\alpha$ is its wavelength. At 4.05 $\mu$m, the line is substantially less affected by interstellar extinction than commonly used near-infrared diagnostics such as Br$\gamma$ at 2.16 $\mu$m. Consequently, Br$\alpha$ may occupy an important niche as an accretion diagnostic for embedded or moderately extincted young stellar objects where shorter-wavelength recombination lines become increasingly attenuated. Because Br$\alpha$ is a lower-order hydrogen recombination transition, it is expected to reach appreciable optical depths under less extreme density and column-density conditions than higher-order infrared H I transitions, including the H I $8-6$ and $10-7$ lines observed with JWST/MIRI \citep[e.g.,][]{Muzerolle2001,Kwan2011}. Br$\alpha$ may therefore probe a larger effective volume of the accretion flow, whereas higher-order transitions can preferentially trace denser regions of the emitting gas. Recent JWST/MIRI studies provide empirical support for the accretion-flow origin of these higher-order lines \citep{Tofflemire2025,Shridharan2026}, with line-ratio modeling indicating hydrogen densities of $\sim10^{10.6}-10^{11.2}$ cm$^{-3}$ in most sources and some objects approaching the optically thin regime \citep{Shridharan2026}. Simultaneous observations of Br$\alpha$, Br$\gamma$, and the mid-infrared hydrogen lines therefore offer a promising avenue for constraining the density, optical depth, and geometry of accretion flows in young stars.

The weak positive correlation between Br$\alpha$ luminosity and stellar bolometric luminosity, together with a similar but statistically non-significant trend with stellar mass, differs from the behavior observed for the CO fundamental emission, which showed little or no dependence on these fundamental stellar properties. These weak trends suggest that Br$\alpha$ emission is influenced not only by the instantaneous accretion luminosity but also, at least indirectly, by the underlying stellar properties that regulate the accretion process itself. In particular, numerous studies have established that the mass accretion rate increases systematically with stellar mass over the PMS regime \citep[e.g.,][and subsequent work]{Manara2012}. More massive stars may therefore sustain larger or denser magnetospheric accretion columns and produce stronger hydrogen recombination emission.

The strong dependence of Br$\alpha$ luminosity on visual extinction and its weaker, suggestive dependence on projected distance from $\theta^1$~Ori~C may also reflect the evolutionary state of the circumstellar environment. Previous studies of the Orion Nebula Cluster have shown that visual extinction decreases statistically with stellar age \citep{Getman2014a}  and that the cluster exhibits a pronounced spatio-temporal age gradient, with younger stars preferentially concentrated toward its central regions \citep{Getman2014b,Getman2018}. If these relations apply to the present sample, then higher $A_V$ and smaller projected distance from $\theta^1$~Ori~C would statistically identify younger systems with more actively accreting inner disks. The observed increase of Br$\alpha$ luminosity with $A_V$ is consistent with this picture and, importantly, is already present in the observed Br$\alpha$ luminosities before extinction correction, indicating that the correlation is not introduced by the dereddening procedure. The radial dependence provides complementary evidence for an evolutionary interpretation, although it should be regarded more cautiously because compact photoionized proplyd emission may preferentially enhance the Br$\alpha$ luminosities of stars projected near $\theta^1$~Ori~C. 

The evolutionary interpretation, however, should not be regarded as unique. Stellar accretion rates are expected to decline over Myr timescales as circumstellar disks evolve, but observations show substantial dispersion in accretion rates at a given age \citep{Manara2023}. Thus, the age gradient inferred from $A_V$ alone does not necessarily provide a quantitative explanation for the observed Br$\alpha$--$A_V$ relation. An alternative, non-exclusive possibility is that $A_V$ also traces aspects of the large-scale gas environment that regulate the supply of material to circumstellar disks. Recent studies have reported spatial correlations in stellar accretion and discussed ongoing or late-stage infall from the surrounding interstellar medium as a possible mechanism linking the large-scale environment to inner-disk accretion \citep{Winter2024}. More broadly, recent work emphasizes that planet-forming disk properties and their evolution may depend on the local environment. This possibility is particularly relevant for the Orion Nebula, where extinction, stellar age, gas density, external irradiation, and projected cluster position are strongly coupled and therefore difficult to disentangle observationally.

Taken together, the Br$\alpha$ correlations are consistent with a gradual decline in inner-disk accretion activity as protoplanetary systems evolve, but they may also reflect environmental regulation of the disk's mass supply. The relatively shallow ATS relation, $\log L_{\rm Br{\alpha}}=28.23 + 0.13 \times A_V$, compared with the much stronger dependence on accretion luminosity, $\log L_{\rm Br\alpha}=29.66+0.70\log(L_{\rm acc}/L_\odot)$, further suggests that $A_V$ is unlikely to be a direct determinant of Br$\alpha$ emission. Instead, extinction may act as a statistical tracer of evolutionary and/or environmental conditions that are themselves coupled to accretion. The weaker dependence on the Spitzer IRAC spectral slope is also consistent with this picture, while suggesting that the gaseous and dusty components of the inner disk may not evolve synchronously. This interpretation parallels our CO results, where warm molecular gas shows substantially stronger dependence on accretion and extinction than on the infrared dust continuum, suggesting that the gas reservoir may evolve more rapidly than the optically thick dust.

Finally, the absence of any correlation between Br$\alpha$ luminosity and stellar X-ray luminosity is not unexpected because magnetospheric accretion and coronal X-ray emission originate from physically distinct processes. The Br$\alpha$ emission detected in the present sample is therefore most naturally interpreted as arising predominantly from accretion columns rather than being directly regulated by the quiescent coronal activity of the host star. Nevertheless, our preliminary thermo-chemical ProDiMo calculations predict that sufficiently energetic X-ray mega-flares can enhance hydrogen recombination emission produced in the irradiated disk atmosphere itself \citep[Paper I;][]{PortillaRevelo2026}. The current Chandra--JWST campaign did not capture such extreme events (Paper I). Future observing programs with larger samples and an increased probability of detecting X-ray mega-flares will therefore provide an important opportunity to determine whether flare-induced disk emission can produce measurable enhancements of the Br$\alpha$ line above the dominant contribution from magnetospheric accretion.

\subsection{CO. Possible astrophysical interpretations} \label{subsection_discussion_CO}

Our analysis shows that the CO fundamental-band luminosities are most strongly correlated with stellar accretion luminosity, while also exhibiting significant positive correlations with visual extinction. The CO--$L_{\rm acc}$ relationship remains evident after excluding stars within $2\arcmin$ of $\theta^1$~Ori~C, although the statistical significance is strongest in the best-sampled CO band, and the CO--Br$\alpha$ correlation persists in the spatially restricted sample. Importantly, the independent NIRSpec+MIRI analysis of PMS stars spanning several star-forming regions recovers similarly strong CO--$L_{\rm acc}$ and CO--$A_V$ correlations, indicating that these relationships are not peculiar to the Orion Nebula Cluster.

Previous high-resolution spectroscopic studies have revealed a dependence of CO fundamental-band luminosity on stellar mass in selected disk populations. In particular, \citet{Banzatti2023} found a strong positive correlation between CO luminosity and stellar mass above approximately 1~M$_{\odot}$ for disks exhibiting double-peaked, predominantly Keplerian CO profiles, but no clear trend for the approximately three times more numerous sources with triangular profiles, interpreted as arising from a combination of disk emission and a low-velocity molecular wind \citep{Banzatti2022}. Thermochemical models also predict increasing infrared molecular emission with stellar mass \citep{Frediani2026}. In contrast, neither our Orion sample nor the combined NIRSpec+MIRI sample exhibits a statistically significant correlation between CO luminosity and stellar mass, despite strong positive correlations with accretion luminosity. This difference may reflect the limited stellar-mass range of our samples, heterogeneous disk evolutionary states, and variations in the CO-emitting area and excitation conditions. Moreover, our moderate-resolution spectra cannot separate the kinematic CO components identified in high-resolution surveys.

Several hypotheses can be considered to explain the NIRSpec CO luminosities measured in our COUP-JWST PMS sample.

{\bf 1. Evolutionary depletion of the warm inner-disk gas}

The simplest interpretation is that the CO fundamental emission traces the amount of warm molecular gas residing within the inner few astronomical units of the disk. If the statistical age-extinction relation previously established for the Orion Nebula Cluster (ONC) \citep{Getman2014b} is applicable to the present sample, then higher-extinction systems are expected to represent younger objects. The observed increase of CO luminosity with $A_V$ would therefore suggest that the warm molecular gas reservoir declines as protoplanetary disks evolve. The simultaneous correlation with accretion luminosity naturally supports this picture. Both stellar accretion and CO fundamental emission require the presence of substantial inner-disk gas, so a gradual depletion of this gas reservoir would reduce both quantities over time.

An intriguing aspect of the results is that the CO luminosities exhibit stronger relationships with $A_V$ and accretion luminosity than with the Spitzer IRAC spectral slope. Indeed, the IRAC slope itself shows weaker correlations with both $A_V$ and $L_{\rm acc}$ (see Appendix~\ref{sec_appendix_multivariate_COUP_analysis}). This may indicate that the warm molecular gas traced by the CO fundamental emission evolves more rapidly than the optically thick dust responsible for the near- and mid-infrared continuum. Such an interpretation is broadly consistent with the growing picture that the gaseous and dusty components of protoplanetary disks do not necessarily evolve synchronously and may disperse on different timescales \citep{Williams2011,Alexander2014,Miotello2017,Ansdell2018}. Recent JWST observations provide an additional illustration of this decoupling: \citet{Romero-Mirza2024} found substantial changes in molecular emission from the AS 209 disk over a decade while the mid-infrared continuum remained nearly unchanged, demonstrating that molecular-gas emission can vary independently of the optically thick dust emission.

{\bf 2. Accretion-driven excitation of CO}

An alternative interpretation is that the observed CO luminosity is determined primarily by the excitation conditions rather than by the total gas mass. Systems with larger accretion luminosities are expected to have stronger accretion-powered UV radiation and accretion-related heating, which can increase the temperature of the inner disk atmosphere and enhance the excitation of the CO fundamental transitions. In addition, the enhanced far-ultraviolet (FUV) radiation produced by the accretion shock may directly excite CO through ultraviolet fluorescence, populating higher vibrational levels that subsequently cascade into the $v=1\rightarrow0$ fundamental transitions observed by JWST/NIRSpec \citep[e.g.,][]{Brittain2007,HeinBertelsen2016}. Under this scenario, the observed CO--$L_{\rm acc}$ correlation would primarily reflect differences in the heating and radiative excitation of the warm molecular gas rather than differences in the total inner-disk gas reservoir. Distinguishing this excitation-driven scenario from an evolutionary depletion of the gas reservoir will require detailed thermo-chemical models that simultaneously predict the gas temperature, chemistry, and emergent CO rovibrational emission.

{\bf 3. A combination of depletion and excitation}

The most realistic explanation is probably that several of these effects operate simultaneously. Disk evolution likely reduces the available warm molecular gas while also decreasing the accretion luminosity and the associated heating of the inner disk atmosphere. Consequently, both the amount of emitting CO and its excitation conditions evolve together, producing the observed correlations.

{\bf 4. An Environmental Origin: Inherited Compactness from Parental Cores}

While the correlation with $A_V$ may point toward evolutionary gas depletion, an alternative is that some low-CO-luminosity systems host intrinsically compact disks with smaller CO-emitting surface areas. Because optically thick CO emission depends more directly on the projected emitting area and excitation temperature than on the total gas mass, compact disks can exhibit weaker integrated CO emission without requiring severe molecular gas depletion. High dust optical depth in the inner disk could further attenuate emission from the warm molecular layer. Such compactness may be inherited from the parental molecular cores: if the age--extinction relation is not the sole driver of $A_V$, lower extinction could also reflect differences in the density or mass of the local birth environment, potentially associated with variations in the mass and angular-momentum content of the parental cores and, consequently, in initial disk sizes and structures. This scenario is consistent with recent high-resolution ALMA observations of faint disks in Lupus \citep{Ricciardi2026}, which reveal a population of intrinsically compact, optically thick disks. If similar systems are present among our low-$A_V$ sources, their weaker CO fundamental-band emission could reflect reduced emitting areas rather than predominantly evolutionary gas depletion. The simultaneous CO--$L_{\rm acc}$ correlation is also compatible with this interpretation if compact disks possess smaller gas reservoirs or less extensive regions capable of sustaining high accretion rates.

\section{Conclusions} \label{sec_conclusions}

As part of our joint Chandra-JWST project investigating the effects of stellar X-ray flares on the chemistry of protoplanetary disks in young stellar members of the Orion Nebula region, this paper presents a population-level spectroscopic analysis of warm inner-disk gas diagnostics using JWST/NIRSpec-MOS observations. Building upon our two-epoch flare variability study in Paper I, we analyzed single-epoch spectra for a substantially expanded sample of low-mass (0.2--1.5~M$_{\odot}$) PMS stars to characterize emission from hydrogen Brackett-$\alpha$ (at 4.05~$\mu$m) and the carbon monoxide fundamental band (4.3--5.2~$\mu$m).

By incorporating censored statistical correlation techniques (Kendall's $\tau$) and Akritas-Theil-Sen (ATS) regression estimators adjusted for multiple testing via the False Discovery Rate ($q_{\rm FDR}$), we investigated how these warm-gas diagnostics relate to fundamental stellar parameters, accretion rates, visual extinction, and cluster environment. 

Our primary results and conclusions are summarized as follows:

{\bf 1) Brackett-$\alpha$ as a Robust Accretion Diagnostic: } Br$\alpha$ line luminosity exhibits an exceptionally strong positive correlation with stellar accretion luminosity ($\tau = 0.58$, $q_{\rm FDR} = 2.8 \times 10^{-5}$). The corresponding ATS regression is $\log(L_{\rm Br\alpha}) = 29.66^{+0.45}_{-0.42} + (0.70^{+0.26}_{-0.22}) \times \log(L_{\rm acc}/L_{\odot})$, where $L_{\rm Br\alpha}$ is in erg s$^{-1}$ and the uncertainties represent 95\% bootstrap confidence intervals. An independent censored ATS regression with the variables exchanged yields $\log(L_{\rm acc}/L_{\odot}) = 4.18^{+1.65}_{-1.82} + (1.19^{+0.29}_{-0.37}) \times \log(L_{\rm Br\alpha}/L_{\odot})$. The latter relation {\it is not an algebraic inversion} of the former. The independently fitted $L_{\rm acc}$--$L_{\rm Br\alpha}$ relation agrees closely with the empirical calibration of \citet{Testi2025}, despite substantial differences in the stellar samples and analysis methods, strengthening the evidence for the broader applicability of Br$\alpha$ as an accretion diagnostic. 

This relationship remains robust after excluding sources projected within $2\arcmin$ of $\theta^1$~Ori~C, showing that the correlation is not driven by the central Trapezium population, where the strongly ionized proplyd environment may introduce systematic uncertainties into $L_{\rm acc}$ estimates. Operating at $4.05~\mu\text{m}$, $\text{Br}\alpha$ is far less attenuated by dust than near-infrared diagnostics like $\text{Br}\gamma$, establishing it as an excellent, extinction-resilient empirical tracer of magnetospheric accretion in PMS stars.

{\bf 2) $\text{Br}\alpha$ Trends and Absence of Quiescent X-ray Coupling:} Br$\alpha$ luminosity correlates positively with visual extinction ($A_V$) and shows a weaker, suggestive correlation with bolometric luminosity ($L_{\rm bol}$), together with a similar but statistically non-significant trend with stellar mass ($M_\star$), potentially reflecting the well-established mass dependence of accretion rates. Conversely, $\text{Br}\alpha$ shows no correlation with quiescent coronal X-ray luminosity ($L_X$), supporting an accretion-related origin rather than a dominant contribution from quiescent coronal activity.

{\bf 3) CO Fundamental Emission Linked to Accretion, Not Stellar Mass:} CO fundamental-band luminosities display highly significant positive correlations with $L_{\text{acc}}$ ($\tau \sim 0.5\text{--}0.6$) and $A_V$ ($\tau \sim 0.35$). Crucially, in contrast to $\text{Br}\alpha$, CO luminosity shows no statistically significant dependence on stellar mass, $L_{\text{bol}}$, or $L_X$. This independence indicates that warm molecular gas in the inner disk is governed primarily by accretion activity and local disk conditions rather than global host-star parameters. 

{\bf 4) Consistency Across Star-Forming Regions:} Combining our 12 complete NIRSpec P-branch spectra with 21 JWST/MIRI spectra of PMS stars in nearby, lower-mass star-forming regions (Taurus, Lupus, Chamaeleon, UCL, and NGC 6357) recovers the same strong CO--$L_{\text{acc}}$ ($\tau = 0.59$, $q = 2.4 \times 10^{-5}$) and CO--$A_V$ ($\tau = 0.34$, $q = 0.008$) correlations. These results indicate that the observed scaling of CO fundamental emission with accretion and extinction is not unique to the high-density ONC environment and may represent a more general property of protoplanetary disks.

{\bf 5) Asynchronous Gas and Dust Evolution:} CO fundamental luminosities correlate poorly with the Spitzer/IRAC spectral energy distribution slope and the IRAC slope itself shows weaker correlations with both $A_V$ and $L_{\rm acc}$. Because $A_V$ serves as a statistical proxy for system age in the ONC, the stronger coupling of CO to $A_V$ and $L_{\text{acc}}$ compared to IRAC slope suggests that the warm molecular gas reservoir in the inner disk may deplete or disperse faster than the optically thick dust continuum.

{\bf 6) Physical Drivers of Inner-Disk CO Emission:} The observed CO trends are consistent with three non-mutually exclusive physical mechanisms. {\bf Evolutionary Gas Depletion:} The decline of CO line strength toward lower $A_V$ and $L_{\rm acc}$ is consistent with progressive depletion or structural evolution of warm inner-disk gas as systems evolve. {\bf Accretion-Driven Excitation:} Higher accretion luminosities can provide stronger accretion-powered UV irradiation and thermal heating, enhancing CO fundamental-band emission through thermal and fluorescent excitation. {\bf Inherited Structural Compactness:} An alternative structural interpretation is that lower-extinction systems may preferentially originate in smaller natal cores and inherit more compact disks.

Detailed thermo-chemical disk modeling, coupled with richer, spatially resolved observations of inner-disk gas geometry, will be essential to disentangle the relative contributions of gas depletion, radiative excitation, and initial structural compactness in shaping the warm molecular atmospheres of young planet-forming disks.

\begin{acknowledgments} We thank the anonymous referee for their careful reading of the manuscript and for helpful comments that improved its clarity. We also thank Gabriel-Dominique Marleau (University of Duisburg-Essen) for bringing relevant literature on Br$\alpha$ as an accretion diagnostic to our attention. This work is supported by STScI grant \#JWST-GO-09292.001-A and SAO Chandra grant GO6-27010X. JF and AB acknowledge support from the Swedish National Space Agency (2022-00154). The JWST/NIRSpec observations analyzed in this work are available from MAST at \dataset[doi:10.17909/3hm9-rw51]{https://doi.org/10.17909/3hm9-rw51}. Support for program \#9292 was provided by NASA through a grant from the Space Telescope Science Institute, which is operated by the Association of Universities for Research in Astronomy, Inc., under NASA contract NAS 5-03127.  This work is based on observations made with the NASA/ESA/CSA James Webb Space Telescope. The data are associated with program 9292 and were obtained from the Mikulski Archive for Space Telescopes (MAST) at the Space Telescope Science Institute, which is operated by the Association of Universities for Research in Astronomy, Inc., under NASA contract NAS 5-03127.
\end{acknowledgments}





%
\facilities{JWST (NIRSpec and MIRI)}

\software{R \citep{RCoreTeam2026}, CTOOL \citep{Pontoppidan2026}}


\clearpage
\appendix

\section{Supplementary Tables}\label{sec_appendix_tables}

This Appendix presents two supplementary tables. (1) Table~\ref{tab_paperII_catalog} provides a catalog of 75 COUP--JWST sources, including source identifiers, coordinates, X-ray and stellar properties, extinction, accretion luminosities, and extinction-corrected H~I Br$\alpha$ and CO line luminosities with their uncertainties or $5\sigma$ upper limits. The full 22-column catalog is available in machine-readable form. (2) Table~\ref{tab_paperII_miri} lists 21 disk-bearing stars observed with JWST/MIRI, providing Gaia DR3 coordinates, distances, extinction, stellar masses, accretion luminosities, and CO luminosities integrated over 4.90--5.11~\micron\ with uncertainties. Literature references for the stellar and accretion properties are also provided.

\begin{deluxetable*}{cl}
\tablecaption{Source Properties and Br$\alpha$/CO Luminosities\label{tab_paperII_catalog}}
\tablehead{\colhead{Column} & \colhead{Description}}
\startdata
\texttt{COUP} & COUP X-ray source identifier (\citealt{Getman2005}) \\
\texttt{JWST} & JWST source identifier (Paper~I) \\
\texttt{RAdeg} & Gaia DR3 ICRS right ascension at epoch 2016.0 (deg) \\
\texttt{DEdeg} & Gaia DR3 ICRS declination at epoch 2016.0 (deg) \\
\texttt{ME} & COUP median X-ray photon energy (keV; \citealt{Getman2005}) \\
\texttt{logLX} & Logarithm of time-averaged COUP X-ray luminosity (erg\,s$^{-1}$; \citealt{Getman2005}) \\
\texttt{IRACslope} & Spitzer/IRAC spectral energy distribution slope (extended catalog of \citealt{Megeath2012}) \\
\texttt{AV} & Visual extinction, $A_V$ (mag; \citealt{Getman2022ApJ}); negative catalog values set to zero \\
\texttt{logLbol} & Logarithm of stellar bolometric luminosity, $\log(L_{\rm bol}/L_\odot)$ (\citealt{Getman2022ApJ}) \\
\texttt{Mass} & Stellar mass, $M_\star$ ($M_\odot$; \citealt{Getman2022ApJ}) \\
\texttt{logLacc} & Logarithm of accretion luminosity, $\log(L_{\rm acc}/L_\odot)$ (\citealt{Manara2012}) \\
\texttt{Rtheta} & Projected angular separation from $\theta^1$ Ori C (\arcsec) \\
\texttt{logLBrA} & $\log(L_{{\rm Br}\alpha}/{\rm erg\,s^{-1}})$, or logarithm of the $5\sigma$ upper limit \\
\texttt{e\_logLBrA} & $1\sigma$ uncertainty in $\log L_{{\rm Br}\alpha}$ (dex) \\
\texttt{logLCO442\_447} & $\log L_{\rm CO}$ over 4.4200--4.4770~$\mu$m (erg\,s$^{-1}$), or $5\sigma$ upper limit \\
\texttt{e\_logLCO442\_447} & $1\sigma$ uncertainty in $\log L_{\rm CO}$ over 4.4200--4.4770~$\mu$m (dex) \\
\texttt{logLCO447\_464} & $\log L_{\rm CO}$ over 4.4770--4.6442~$\mu$m (erg\,s$^{-1}$), or $5\sigma$ upper limit \\
\texttt{e\_logLCO447\_464} & $1\sigma$ uncertainty in $\log L_{\rm CO}$ over 4.4770--4.6442~$\mu$m (dex) \\
\texttt{logLCO468\_500} & $\log L_{\rm CO}$ over 4.6780--5.0000~$\mu$m (erg\,s$^{-1}$), or $5\sigma$ upper limit \\
\texttt{e\_logLCO468\_500} & $1\sigma$ uncertainty in $\log L_{\rm CO}$ over 4.6780--5.0000~$\mu$m (dex) \\
\texttt{logLCO490\_511} & $\log L_{\rm CO}$ over 4.9000--5.1100~$\mu$m (erg\,s$^{-1}$); 12 selected sources only \\
\texttt{e\_logLCO490\_511} & $1\sigma$ uncertainty in $\log L_{\rm CO}$ over 4.9000--5.1100~$\mu$m (dex) \\
\enddata
\tablecomments{The full 75-source catalog combines 73 Br$\alpha$ and 46 CO sources, of which 44 are common to both samples. The full 75-source, 22-column table is provided in machine-readable form. All line luminosities are corrected for flux recovery and extinction. For non-detections, the luminosity column gives $\log_{10}(5\sigma_L)$ and its corresponding uncertainty column is marked as unavailable. When both luminosity and uncertainty entries are unavailable, the measurement is missing because of incomplete spectral coverage or exclusion from the relevant subsample. Logarithmic uncertainties for detections are calculated as $\sigma_L/(L\ln 10)$ in dex. The 4.90--5.11~$\mu$m CO band is reported only for the 12 Orion NIRSpec sources included in the NIRSpec+MIRI comparison.}
\end{deluxetable*}

\begin{deluxetable*}{lrrrrrrrrl}
\tabletypesize{\footnotesize}
\tablecaption{Stellar Properties and MIRI CO Luminosities\label{tab_paperII_miri}}
\tablewidth{0pt}
\tablehead{
\colhead{Source} & \colhead{RA} & \colhead{Dec} & \colhead{$d$} & \colhead{$A_V$} & \colhead{$M_\star$} & \colhead{$\log L_{\rm acc}$} & \colhead{$\log L_{\rm CO}$} & \colhead{$\sigma_{\log L}$} & \colhead{Refs} \\
\colhead{} & \colhead{deg} & \colhead{deg} & \colhead{pc} & \colhead{mag} & \colhead{$M_\odot$} & \colhead{$L_\odot$} & \colhead{erg\,s$^{-1}$} & \colhead{dex} & \colhead{} \\
\colhead{(1)} & \colhead{(2)} & \colhead{(3)} & \colhead{(4)} & \colhead{(5)} & \colhead{(6)} & \colhead{(7)} & \colhead{(8)} & \colhead{(9)} & \colhead{(10)} \\
}
\startdata
CX Tau & 63.6994666 & 26.8029596 & 126.73 & 0.25 & 0.33 & -3.07 & 28.96 & 0.17 & 12,6,13 \\
CY Tau & 64.3905803 & 28.3462226 & 126.33 & 0.35 & 0.37 & -2.00 & 29.33 & 0.10 & 12,6,3 \\
GW Lup & 236.6862963 & -34.5100142 & 155.20 & 0.80 & 0.41 & -2.62 & 29.16 & 0.16 & 4,6,13 \\
EX Lup & 240.7728221 & -40.3071631 & 154.72 & 1.10 & 0.60 & -0.70 & 29.35 & 0.24 & 1,15,1 \\
BP Tau & 64.8160201 & 29.1073641 & 127.40 & 0.45 & 0.47 & -1.13 & 29.89 & 0.03 & 12,6,13 \\
AA Tau & 68.7309435 & 24.4813063 & 134.67 & 0.40 & 0.47 & -1.60 & 29.81 & 0.06 & 12,6,5 \\
DN Tau & 68.8641029 & 24.2496046 & 128.60 & 0.55 & 0.50 & -2.10 & 29.01 & 0.19 & 12,6,3 \\
IQ Tau & 67.4648501 & 26.1123663 & 131.51 & 0.85 & 0.50 & -1.21 & 29.49 & 0.15 & 12,2,13 \\
DF Tau & 66.7616492 & 25.7061194 & 176.45 & 0.10 & 0.61 & -0.55 & 30.70 & 0.02 & 12,6,13 \\
GK Tau & 68.3940490 & 24.3515353 & 129.14 & 1.35 & 0.67 & -1.50 & 29.85 & 0.04 & 12,2,13 \\
Sz 98 & 242.0936679 & -39.0796721 & 156.27 & 1.00 & 0.67 & -0.50 & 29.38 & 0.19 & 1,6,1 \\
VW Cha & 167.0058162 & -77.7079357 & 188.16 & 1.90 & 0.75 & -0.78 & 31.09 & 0.03 & 9,6,9 \\
CI Tau & 68.4667698 & 22.8416169 & 160.32 & 1.90 & 0.71 & -0.20 & 30.13 & 0.06 & 12,2,13 \\
PDS 70 & 212.0421347 & -41.3980439 & 112.39 & 0.05 & 0.76 & -3.34 & 28.76 & 0.27 & 10,10,13 \\
V1094 Sco & 242.1506809 & -39.3841214 & 154.76 & 1.75 & 0.83 & -2.43 & 29.01 & 0.23 & 14,6,13 \\
SY Cha & 164.1261420 & -77.1942659 & 180.73 & 0.83 & 0.85 & -2.09 & 29.03 & 0.31 & 9,6,13 \\
DL Tau & 68.4128654 & 25.3438348 & 159.94 & 1.80 & 0.89 & -0.07 & 30.23 & 0.03 & 8,6,13 \\
LkCa 15 & 69.8241806 & 22.3508638 & 157.19 & 0.30 & 0.91 & -1.89 & 29.10 & 0.20 & 12,6,3 \\
DR Tau & 71.7759207 & 16.9784977 & 192.96 & 0.45 & 1.11 & 0.58 & 30.95 & 0.06 & 8,6,13 \\
IM Lup & 239.0382932 & -37.9351405 & 155.82 & 0.40 & 0.95 & -1.10 & 30.00 & 0.06 & 7,6,1 \\
XUE 1 & 261.1670845 & -34.2070465 & 1690.00 & 9.20 & 1.20 & NA & 30.52 & 0.07 & 11,11,NA \\
\enddata
\tablecomments{The 21 MIRI sources are shown. Columns 2--3 give Gaia DR3 ICRS coordinates at epoch 2016.0. Column 4 gives $d=1000/\varpi$ (pc), with Gaia parallax $\varpi$ in mas, except for XUE~1, for which $d=1690$~pc is adopted from \citet{Ramirez-Tannus2025}. Columns 5--7 give extinction, stellar mass, and $\log(L_{\rm acc}/L_\odot)$. Columns 8--9 give $\log(L_{\rm CO}/[\mathrm{erg\,s}^{-1}])$ integrated over 4.90--5.11~\micron\ and its $1\sigma$ uncertainty in dex. Column 10 lists the literature references for $A_V$, $M_\star$, and $L_{\rm acc}$, respectively; NA indicates unavailable information. References: 1 = \citet{Alcala2017}; 2 = \citet{Banzatti2015}; 3 = \citet{Gangi2022}; 4 = \citet{Garufi2022}; 5 = \citet{Hartmann1998}; 6 = \citet{Henning2024}; 7 = \citet{Herczeg2014}; 8 = \citet{Long2019}; 9 = \citet{Manara2016}; 10 = \citet{Muller2018}; 11 = \citet{Ramirez-Tannus2025}; 12 = \citet{SullivanKraus2022}; 13 = \citet{Tofflemire2025}; 14 = \citet{Tsukagoshi2011}; 15 = \citet{Woitke2024}.}
\end{deluxetable*}

\clearpage

\section{Multivariate Analysis of Stellar and Environmental Properties in the Orion Nebula} \label{sec_appendix_multivariate_COUP_analysis}

To place the Br$\alpha$ and CO correlation analyses presented in the main text into a broader context, we performed a multivariate analysis of stellar and environmental properties for a large sample of Orion Nebula COUP PMS stars \citep{Getman2005}. The parent sample was restricted to systems with $A_V<10$ mag and $M<2$~M$_{\odot}$, making it more representative of the COUP--JWST sample by excluding most highly embedded protostars in OMC-1 and the more massive stars in the nebula.

The analyzed quantities include the X-ray median photon energy \citep[$ME$; a surrogate for both the coronal plasma temperature and the intervening gas column density;][]{Getman2010}, X-ray luminosity ($L_X$), effective temperature ($T_{\rm eff}$), bolometric luminosity ($L_{\rm bol}$), stellar mass ($M$), accretion luminosity ($L_{\rm acc}$), visual extinction ($A_V$), Spitzer IRAC spectral slope ($\alpha_{\rm IRAC}$), and projected angular distance from $\theta^1$~Ori~C. The stellar parameters ($ME$, $L_X$, $T_{\rm eff}$, $L_{\rm bol}$, $M$, and $A_V$) are adopted from Table~6 of \citet{Getman2022ApJ}, IRAC slopes from the extended catalog of \citet{Megeath2012}, and accretion luminosities from \citet{Manara2012}.

\begin{figure*}[!t]
\centering
\includegraphics[width=0.90\textwidth]{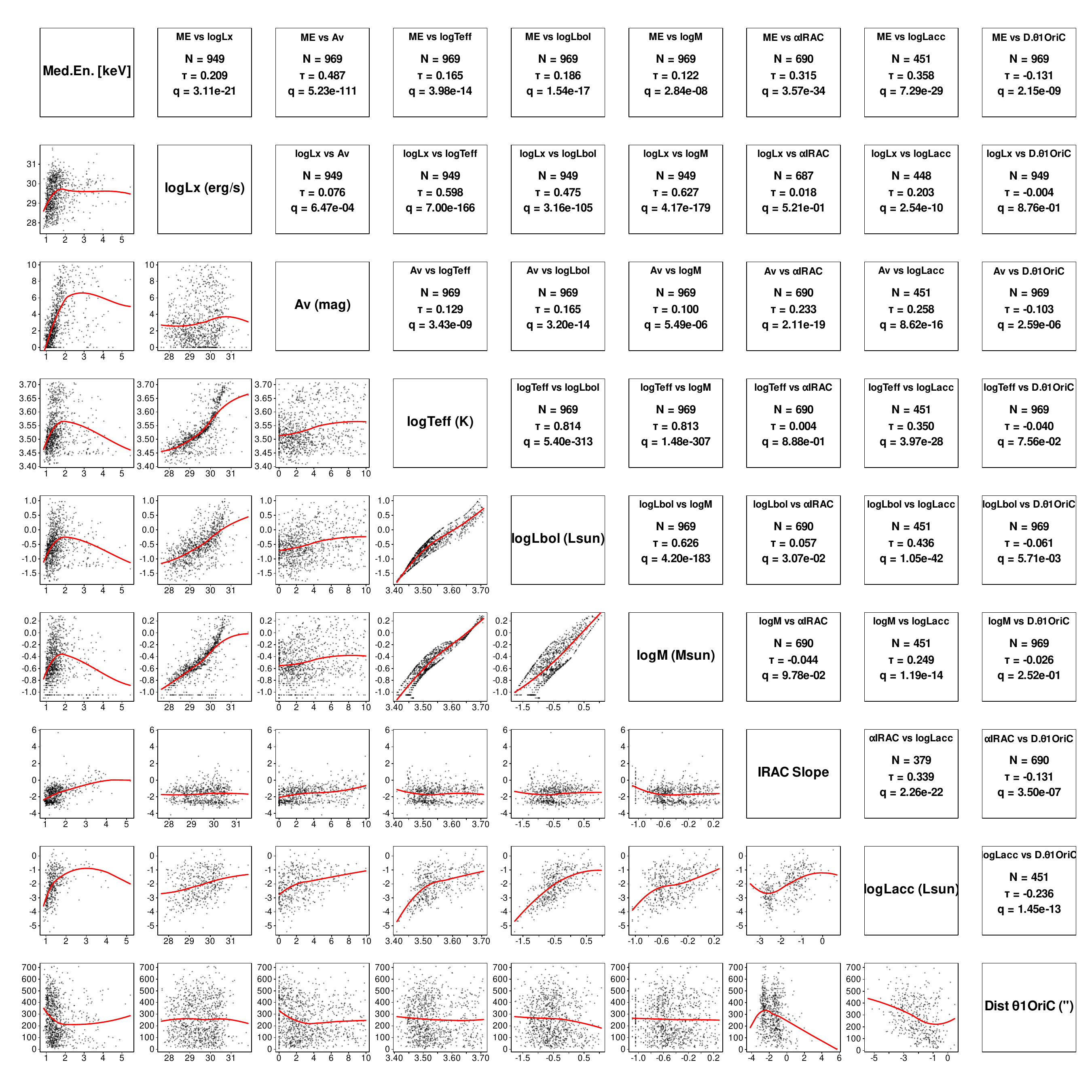}
\caption{Pairs plot showing the bivariate relationships among the stellar and environmental properties of a large sample of PMS stars in the Orion Nebula. Bivariate scatter plots with LOESS \citep{Cleveland1979} fits (red curves) are shown below the diagonal, while the panels above the diagonal report the number of systems included in each comparison, Kendall's $\tau$, and the false-discovery-rate (FDR)-corrected ($q$) values \citep{BenjaminiHochberg1995} (i.e., adjusted for multiple testing).} \label{fig_appendix_multivar}
\end{figure*}

Figure~\ref{fig_appendix_multivar} summarizes the results of the pairwise Kendall's $\tau$ rank correlation analysis. The reported significance levels correspond to false-discovery-rate (FDR) corrected $q$-values. Owing to the large sample sizes ($N\sim400$--1000), the statistical power is high: 78\% of the tested relationships remain highly significant after FDR correction ($q<0.001$), even when the correlation strengths are modest. The measured Kendall's $\tau$ coefficients span the full range from negligible ($|\tau|<0.1$) to extremely strong ($|\tau|>0.8$).

Several trends are particularly relevant for interpreting the Br$\alpha$ and CO results presented in this paper.

{\bf (1) Evolutionary tracers of circumstellar material.} X-ray median photon energy, visual extinction, IRAC spectral slope, and accretion luminosity exhibit highly significant positive correlations among each other, with Kendall's $\tau \simeq 0.23$--0.49. These quantities therefore appear to trace, at least in part, the evolutionary state of PMS star--disk systems, with younger systems tending to exhibit larger gas and dust columns, redder infrared SEDs, and higher accretion rates.

{\bf (2) Fundamental stellar properties.} Stellar mass, effective temperature, bolometric luminosity, X-ray luminosity, and accretion luminosity are all positively correlated among each other ($\tau \simeq 0.20$--0.81). These relationships primarily reflect the underlying dependence of stellar luminosity, temperature, magnetic activity, and accretion on stellar mass, together with the evolutionary properties of young ($\lesssim 1$--2 Myr) fully convective PMS stars.

{\bf (3) Weak coupling between stellar and disk evolutionary indicators.} Fundamental stellar quantities ($L_X$, $T_{\rm eff}$, $L_{\rm bol}$, and $M$) show weak or no correlations with the disk evolutionary indicators $\alpha_{\rm IRAC}$ and $A_V$. The moderate correlation between $L_X$ and $ME$ largely reflects the dual sensitivity of $ME$ to both interstellar absorption and intrinsic coronal plasma temperature.

{\bf (4) Distance from $\theta^1$~Ori~C.} The projected distance from $\theta^1$~Ori~C exhibits weak negative correlations with $ME$, $A_V$, and $\alpha_{\rm IRAC}$, but little or no correlation with the fundamental stellar parameters ($L_X$, $T_{\rm eff}$, $L_{\rm bol}$, and $M$). These weak trends are consistent with previous evidence for spatial age gradients across the ONC \citep{Getman2014b,Getman2018}, suggesting that distance from $\theta^1$~Ori~C may serve as a secondary indicator of evolutionary state. In contrast, $L_{\rm acc}$ shows a moderate negative correlation with projected distance. One possible explanation is an observational bias in the accretion luminosities derived from U-band excess and H$\alpha$ measurements \citep{Manara2012}, since stars projected closer to $\theta^1$~Ori~C experience stronger nebular Balmer continuum and H$\alpha$ background, as well as brighter photoevaporative proplyd emission.

{\bf (5) Implications for the Br$\alpha$ and CO analyses.} The large statistical power of this multivariate analysis demonstrates that intrinsically weak correlations may require samples of several hundred objects to achieve statistical significance. Consequently, the absence of significant correlations in our Br$\alpha$ and CO analyses ($N\sim30$--70) should not necessarily be interpreted as evidence for the absence of an intrinsic relationship. For example, although the present NIRSpec data do not reveal a statistically significant correlation between CO luminosity and IRAC spectral slope ($q>0.05$), the measured $\tau \sim 0.2$--0.3 suggests a potentially weak underlying trend. If both quantities partially trace the evolution of inner-disk gas and dust, larger future samples may establish this relationship with statistical significance.

\clearpage

\bibliography{sample702}{}
\bibliographystyle{aasjournalv7.1}



\end{document}